\documentclass[11pt,titlepage]{article} 
\usepackage[numbers,square,sort&compress]{natbib}
\usepackage{amsfonts}
\usepackage{amsmath}
\usepackage{amssymb}
\usepackage{url}

\usepackage{caption}
\usepackage{graphicx,rotating}

\usepackage{setspace}
\usepackage[table]{xcolor}
\begin{document}
\title{Inference of population splits and mixtures from genome-wide allele frequency data}
 \author{Joseph K. Pickrell$^{1,3,\dagger}$, Jonathan K. Pritchard$^{1,2,\dagger}$\\ 
 \\
 \small $^1$ Department of Human Genetics and\\
 \small $^2$ Howard Hughes Medical Institute, University of Chicago\\
 \small $^3$ Current address: Department of Genetics, Harvard Medical School\\
 \small $\dagger$ To whom correspondence should be addressed: joseph\_pickrell@hms.harvard.edu, pritch@uchicago.edu
 }

\maketitle
\pagebreak
\begin{abstract}
Many aspects of the historical relationships between populations in a species are reflected in genetic data. Inferring these relationships from genetic data, however, remains a challenging task. In this paper, we present a statistical model for inferring the patterns of population splits and mixtures in multiple populations. In this model, the sampled populations in a species are related to their common ancestor through a graph of ancestral populations. Using genome-wide allele frequency data and a Gaussian approximation to genetic drift, we infer the structure of this graph. We applied this method to a set of 55 human populations and a set of 82 dog breeds and wild canids. In both species, we show that a simple bifurcating tree does not fully describe the data; in contrast, we infer many migration events. While some of the migration events that we find have been detected previously, many have not. For example, in the human data we infer that Cambodians trace approximately 16\% of their ancestry to a population ancestral to other extant East Asian populations. In the dog data, we infer that both the boxer and basenji trace a considerable fraction of their ancestry (9\% and 25\%, respectively) to wolves subsequent to domestication, and that East Asian toy breeds (the Shih Tzu and the Pekingese) result from admixture between modern toy breeds and ``ancient" Asian breeds. Software implementing the model described here, called \emph{TreeMix}, is available at \url{http://treemix.googlecode.com}. 

\end{abstract}

\section*{Author Summary}
With modern genotyping technology, it is now possible to obtain large amounts of genetic data from many populations in a species. An important question that can be addressed with these data is: what is the history of these populations? There is a long history in population genetics of inferring the relationships among population as a bifurcating tree, analogous to phylogenetic trees for representing the evolution of species. However, it has long been recognized that, since populations from the same species exchange genes, simple bifurcating trees may be an incorrect representation of population histories. We have developed a method to address this issue, using a model which allows for both population splits and gene flow. In application to humans, we show that we are able to identify a number of both previously known and unknown episodes of gene flow in history, including gene flow into Cambodia of a population only distantly related to modern East Asia. In application to dogs, we show that the boxer and basenji breeds have a considerable component of ancestry from grey wolves subsequent to domestication. 

\section*{Introduction}

The extant populations in a species result from an often-complex demographic history, involving population splits, gene flow, and changes in population size. It has long been recognized that genetic data can be used to learn about this history \citep{Cavalli-Sforza:1967fk, Felsenstein:1982vn, Cann:1987ly}. In humans, early approaches to inferring history from genetics were limited to using a relatively small number of blood group or other protein polymorphisms \citep{Cavalli-Sforza:1967fk, Nei:1974ve, Nei:1993qf, Cavalli-Sforza:1988bh}. These types of studies were then superseded by analyses of DNA markers, which have progressed from single marker studies \citep{Cann:1987ly} to studies involving hundreds of thousands of markers \citep{Li:2008rt}. It is now feasible to collect genome-wide genetic data in any species; to a large extent it is no longer the data collection, but rather the statistical models used for analysis, that limit the historical insight possible.   

There are many statistical approaches to demographic inference from genetic data. One approach is to develop an explicit population genetic model for the history of a set of populations, framed in terms of the effective population sizes of the populations, the times of population splits, the times of demographic events (such as population bottlenecks), and other relevant parameters. The values of these parameters can then be learned from the data using a variety of techniques, often involving simulation \citep{Pritchard:1999kx, Beaumont:2002vn, Schaffner:2005ys,Gronau:2011fk, Gutenkunst:2009zr, Hey:2004fk, Beerli:1999uq, Becquet:2007dq, Kubatko:2009kx}. These approaches have the advantage of allowing flexible modeling of a wide variety of demographic scenarios, but the disadvantage that they can only be applied to one or a few populations at a time.  

Another type of approach to learning about population history uses methods that summarize the major components of genetic variation in a sample by clustering or principal components analysis \citep{Menozzi:1978uq, Pritchard:2000zr, Patterson:2006ve, Lawson:2012fk}. Although these methods do not model history explicitly, the inferred components can often be interpreted \emph{post hoc} as representing historical populations, and individuals or populations that are mixtures of different components as evidence of admixture between these populations (e.g.,  \citep{Menozzi:1978uq, Rosenberg:2002fr, Liti:2009fk, Vonholdt:2010uq}). However, these methods are not directly informative about history; indeed, the relationship between the major components of genetic variation and true underlying demography is not always intuitive \citep{Francois:2010vn, McVean:2009zr, Novembre:2008nx}. 

A different class of approaches focuses on the relationships between populations, by representing a set of populations as a bifurcating tree  \citep{Cavalli-Sforza:1967fk,Saitou:1987uq, Felsenstein:1973kx, RoyChoudhury:2008vn, Felsenstein_1981, Siren:2011fk, Nielsen_Mountain_Huelsenbeck_Slatkin_1998}. In these models, the details of the demographic histories of the population are absorbed into the branch lengths of the tree \citep{Cavalli-Sforza:1967fk, Nicholson:2002gf}. This approach has the advantage of being applicable to large numbers of populations; however, a major caveat when modeling the history of populations as a tree is that gene flow violates the assumptions of the model \citep{Cavalli-Sforza:1973mz, Cavalli-Sforza:1975gf, Felsenstein:1982vn}. It is often difficult to know, \emph{a priori}, how well the history fits a simple bifurcating tree. Explicit tests for the violation of a tree model have been developed \citep{Keinan:2007kx, Reich:2009fk, Durand:2011kx, Green:2010vn, Cavalli-Sforza:1975gf, Lathrop:1982uq}. These tests have been used, most notably, to infer the existence of gene flow between modern and archaic humans \citep{Reich:2010uq, Green:2010vn, Reich:2011uq}, as well as between diverged modern human populations \citep{Reich:2009fk, Moorjani:2011ys, Rasmussen:2011fk}. 

In this paper, we present a unified statistical framework for building population trees and testing for the presence of gene flow between diverged populations. In this framework, the relationship between populations is represented as a graph, allowing us to model both population splits and gene flow. Graph-based models are of growing interest in phylogenetics  \citep{phylogenetic.networks, Huson:2006fk}, but have been rarely used in population genetics (with some exceptions \citep{Lathrop:1982uq, Dyer:2004uq, Reich:2009fk}). 

\section*{Results}
The starting point for our model was first proposed by Cavalli-Sforza and Edwards \citep{Cavalli-Sforza:1967fk}, and we draw additionally on related models by Nicholson et al. \citep{Nicholson:2002gf} and Coop et al. \citep{Coop:2010ly}. Our goal is to provide a statistical framework for inferring population networks that is motivated by an explicit population genetic model, but sufficiently abstract to be computationally feasible for genome-wide data from many populations (say, 10-100). Our primary aim is to represent the topology of relationships between populations, rather than the precise times of demographic events. 

Our approach to this problem is to first build a maximum likelihood tree of populations. We then identify populations that are poor fits to the tree model, and model migration events involving these populations. Below, we first describe this approach in an idealized setting, and then describe the modifications necessary for implementation in practice. 
\subsection*{Model}




\begin{figure}
\begin{center}
\includegraphics[scale = 0.8]{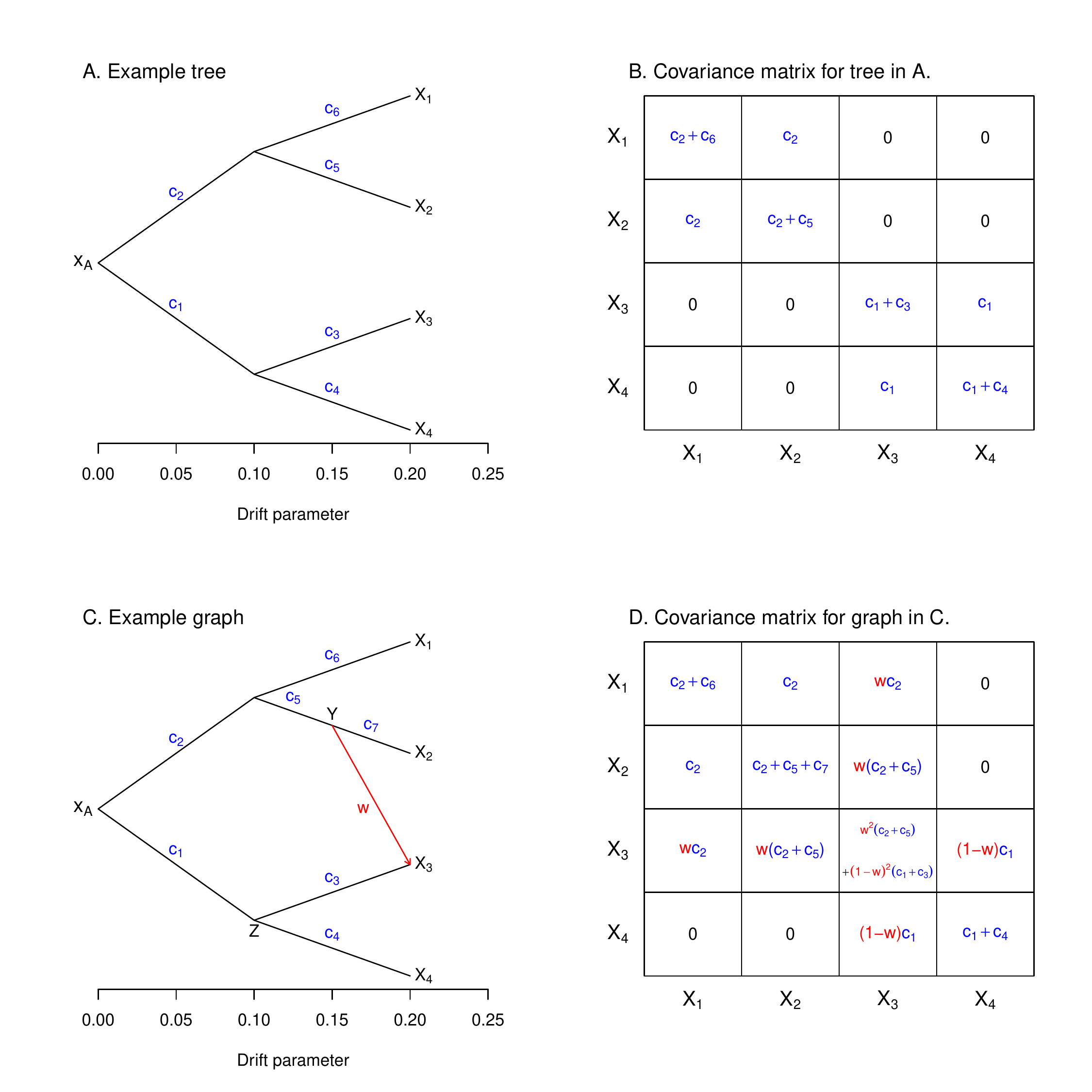}

\caption{\textbf{: Simple examples. A.} An example tree. \textbf{B.} The covariance matrix implied by the tree structure in A. Note that the covariance here is with respect to the allele frequency at the root, and that each entry has been divided by $x_A[1-x_A]$ to simplify the presentation. \textbf{C.} An example graph. The migration edge is colored red. Parental populations for population 3 are labeled $Y$ and $Z$; see the main text for details. \textbf{D.} The covariance matrix implied by the graph in C; again, each entry has been divided by $x_A[1-x_A]$. The migration terms are in red, and the non-migration terms are in blue. }\label{ex_fig}

\end{center}
\end{figure}

In the most simple case, consider a single SNP, and let the allele frequency of one of the alleles at this SNP in an ancestral population be $x_A$ (We use a lowercase $x$ to denote that this is a parameter rather than a random variable. We initially consider distributions conditional on $x_A$). Now consider a descendant population $B$. We model $X_B$, the allele frequency of the SNP in population $B$, as:
\begin{equation} 
X_B  = x_A + \epsilon_B
\end{equation}
\noindent with
\begin{equation} \label{normal_model}
\epsilon_B \sim N(0, c_B x_A [1-x_A])
\end{equation}

\noindent where $c_B$ is a factor that corresponds to the amount of genetic drift that has occurred between the ancestral population and $B$. This Gaussian model was first introduced by Cavalli-Sforza and Edwards \citep{Cavalli-Sforza:1967fk}, and the motivation for this model is outlined in Nicholson et al. \citep{Nicholson:2002gf}. Briefly, if the amount of genetic drift between the two populations is small (at most on a timescale of the same order as the effective population size), then the diffusion approximation to a Wright-Fisher model of genetic drift leads to Equation \ref{normal_model} with $c_B \approx \frac{t}{2N_e}$, where $t$ is the number of generations separating the two populations, and $N_e$ is the effective population size \citep{Nicholson:2002gf}. We do not model the boundaries of the allele frequencies at zero and one, nor do we consider new mutations. This means that this model will be most accurate for alleles that were at intermediate frequency in the ancestral population. 

Now consider a descendant population of $B$; let us call this population $C$, and the allele frequency in the population $X_C$. Using the same model:
\begin{align}
X_C & = X_B + \epsilon_C \\
&= x_A + \epsilon_B + \epsilon_C
\end{align}

\noindent where 
\begin{equation}
\epsilon_C \sim N(0, c_C X_B[1-X_B]).
\end{equation}

\noindent We can write down the expectation and variance of $X_C$ as:
\begin{align}
E[X_C] &= E[ x_A+ \epsilon_B + \epsilon_C] \\
&= x_A
\end{align}

\noindent and:
\begin{align}
Var(X_C) &= Var( x_A+ \epsilon_B + \epsilon_C) \\
&= Var(\epsilon_B) + Var(\epsilon_C) + 2Cov(\epsilon_B, \epsilon_C).
\end{align}

\noindent We then assume that the amount of genetic drift between all the populations is small. This implies that $X_B [1-X_B]$ is well-approximated by $x_A [1-x_A]$ in Equation 5, and hence  the amount of genetic drift between $A$ and $B$ is approximately independent of the amount of genetic drift between $B$ and $C$ \citep{Cavalli-Sforza:1975gf}. With these simplifications:
\begin{align}
Var(X_C) &\approx  Var(\epsilon_B)+Var(\epsilon_C)\\
&\approx (c_B+c_C) x_A[1-x_A].
\end{align}

\noindent We thus have a model for $X_C$, conditional on $x_A$:
\begin{equation}
X_C \sim N(x_A, (c_B+c_C) x_A[1-x_A]).
\end{equation}

\paragraph{Multiple populations.}
Now consider a set of four populations, all related to an ancestral population by a tree, as depicted in Figure \ref{ex_fig}A. Let the allele frequencies in the four populations be denoted $X_1$, $X_2$, $X_3$, and $X_4$, respectively, and the vector of all four frequencies be  $\vec X$. We want to write down a joint distribution for  $\vec X$ given the tree. We start by writing down the covariance between any two populations with respect to the ancestral allele frequency (i.e. $Cov(X_i, X_j) = E[ (X_i- x_A)(X_j-x_A)]$). This is simply the variance of the common ancestor of the two populations:
\begin{align}
Cov(X_1, X_2) &= c_2 x_A[1-x_A]\\
Cov(X_3, X_4) &= c_1 x_A[1-x_A]\\
Cov(X_1, X_3) &=0
\end{align} 
\noindent and so on (Figure \ref{ex_fig}B).

Let us use $\bf V$ to denote the variance-covariance matrix of allele frequencies between populations implied by the tree. Now, if we knew the value of $x_A$, we could model $\vec X$ as:

\begin{equation}\label{eq_mvn1}
\vec X \sim MVN(\vec x_A, \bf V)
\end{equation}

\noindent where $\vec x_A = [x_A, x_A, x_A, x_A]$ and $MVN$ denotes the multivariate normal distribution. The covariance matrix with respect to the root, $\bf V$, is distinct from the sample covariance matrix; we return to this distinction later. This multivariate normal model was proposed by Felsenstein \citep{Felsenstein:1973kx}. Additionally, a multivariate normal model was used by Coop et al. \citep{Coop:2010ly} and Weir and Hill \citep{Weir:2002uq}, although these authors did not explicitly model the variance-covariance matrix in terms of a tree.

\paragraph{Modeling migration.} To extend this framework to include migration, we allow populations to have ancestry from multiple parental populations \citep{Lathrop:1982uq, Cavalli-Sforza:1975gf}. The contribution of each parental population is weighted; if we assume admixture occurs in a single generation, these weights are related to (though not necessarily equivalent to) the fraction of alleles in the descendant population that originated in each parental population. Consider population 3 in Figure \ref{ex_fig}C (recall that the allele frequency in this population is $X_3$). We have labeled the two parental populations $Y$ and $Z$; let the allele frequencies in these populations be $X_Y$ and $X_Z$, respectively. We model $X_3$ as:

\begin{equation}
X_3 = wX_Y + (1-w)( X_Z + \epsilon_3)
\end{equation}
\noindent where $\epsilon_3 \sim N(0, c_3 x_A[1-x_A])$. Note that there is some question as to how to weight $\epsilon_3$, the genetic drift specific to population 3. In reality, it comes from three sources: drift since $Y$ but before the population mixture, drift since $Z$ but before the population mixture, and drift since the mixture. These three components should have weights $w$, $1-w$, and 1, respectively. However, the three components are not all separately identifiable. For ease of computation, we estimate only a single drift parameter in this situation, and weight it by $1-w$ (Supplementary Information). 

Additionally, there is a choice of whether the edge from $Z$ or the edge from $Y$ should be considered the ``migration" edge; these two possibilities not identifiable in the model. In practice, we set the edge with the largest weight as the non-migration edge, and the other edge (or edges) as the ``migration" edge(s).

With these simplifications, the variance of $X_3$ can be written in the mixture case as:
\begin{align}
Var(X_3) &= Var(wX_Y + (1-w)(X_Z+\epsilon_3))\\
&= w^2Var(X_Y) + (1-w)^2 [Var(X_Z) + Var(\epsilon_3)]+ 2w(1-w)Cov(X_Y, X_Z) 
\end{align}

\noindent We can now consider multiple populations related by a graph instead of a tree (Figure \ref{ex_fig}C). The variance-covariance matrix $\bf V$ can be filled in as before, but now including terms for migration (Figure \ref{ex_fig}D). This model can be written in terms of a directed acyclic graph (the lack of cycles follows from the fact that no population can contribute genetic material to its own ancestor), where the $c$ parameters correspond to edge lengths (Supplementary Material). For subsets of up to four populations, this model is closely related to the ``$f-$ statistics" used as tests for treeness by  Reich et al. \citep{Reich:2009fk} (Supplementary Material).

\paragraph{Normalization.}
As described above, $\bf V$ depends on the ancestral allele frequency $x_A$. This means that the true variance-covariance matrix will differ by a scaling factor between SNPs with different values of $x_A$. In much work on this type of model, investigators have normalized allele frequencies to account for this. One potential normalization is the arcsine square-root transformation \citep{Cavalli-Sforza:1967fk}. However, a drawback to this normalization is that it is non-linear; the expected value of the allele frequency in the descendant populations is no longer $x_A$, but pushed towards the boundaries, which could induce spurious correlations between the most drifted populations \citep{citeulike:584535}. Another plausible transformation would be to scale all allele frequencies by $\hat \mu(1-\hat \mu)$, where $\hat \mu$ is the mean allele frequency across populations \citep{Nicholson:2002gf, Patterson:2006ve}. Both of these transformations increase the influence of polymorphisms that were rare in the ancestral population. However, these are precisely the loci where the approximation of our model to a true population genetics model is most likely to break down. For these reasons, we choose to work directly with untransformed allele frequencies. 

\paragraph{Properties of the sample covariance.}
In practice, the multivariate normal model in Equation \ref{eq_mvn1} is impractical because we do not know the ancestral values of allele frequencies, but instead only the values in sampled descendant populations. This means that $\bf{V}$ cannot be estimated directly from data. However, consider instead the \emph{sample} covariance matrix $\bf W$, where ${\bf W}_{ij}= E[ (X_i - \hat \mu)(X_j -\hat \mu)]$, where $\hat \mu = \frac{\sum_{i = 1}^m X_i}{m}$, $m$ is the number of populations, and $X_i$ and $X_j$ are the sample allele frequencies in populations $i$ and $j$. $\bf W$ is closely related to $\bf V$ as follows:

\begin{align}
{\bf W}_{ij} &= E[ (X_i - \hat \mu)( X_j - \hat \mu)]\\
&= E[ (X_i - x_A - \hat \mu +x_A)(  X_j - x_A - \hat \mu + x_A)]\\
&= E[ (X_i- x_A)(X_j - x_A) - (X_i -x_A)( \hat \mu - x_A) - (X_j -x_A)(\hat \mu - x_A) +(\hat \mu - x_A)^2]\\
&= {\bf V}_{ij} - \frac{1}{m} \sum_{k = 1}^m {\bf V}_{ik} - \frac{1}{m} \sum_{k = 1}^m {\bf V}_{jk} + \frac{1}{m^2} \sum_{k = 1}^m \sum_{k' = 1}^m {\bf V}_{kk'}.
\end{align}

\noindent In the following section, we will describe how we perform inference based on the sample covariance matrix $\bf W$. 

\paragraph{Finite samples and multiple (potentially correlated) SNPs.}
Now assume that we have genotyped $n$ SNPs in each of $m$ populations. Let the sample allele frequency at SNP $k$ in population $i$ be $\hat X_{ik}$. We can estimate the sample covariance matrix ${\bf \hat W}$:

\begin{equation}
{\bf \hat W}_{ij} = \frac{ \sum_{k = 1}^n [( \hat X_{ik} - \hat \mu_k)( \hat X_{jk} - \hat \mu_k)]}{n}  
\end{equation}

\noindent where $\hat \mu_k = \frac{1}{m} \sum_{i = 1}^m \hat X_{ik}$. The fact that in practice we have finite samples from a population has two effects on this matrix. First, sampling leads to a biased estimation of the terms on the diagonal, since sampling can be thought of as adding an amount of ``drift" to each population (as well as smaller effect on the off-diagonal terms; see Supplementary Material for details). Additionally, it leads to some uncertainty in the off-diagonal terms. To account for the biased diagonal terms, in practice we calculate a corrected version of $\bf \hat W$ that removes this bias (Supplementary Material). To account for uncertainty in the values of this matrix, we use a block resampling approach (see below). Finally, with multiple SNPs, we are working with SNPs with many different values of $x_A$. In this case, the $x_A[1-x_A]$ terms described above can be thought of as $\overline {x_A[1-x_A]}$; i.e., the mean across SNPs of $x_A[1-x_A]$. 

We now want to write down a likelihood for $\bf \hat W$ given $\bf W$. One possibility would be to use the Wishart distribution, since the sample covariance matrix of multivariate normal random variables has this form. However, computation of the Wishart density involves computationally-intensive matrix inversion, so we took an alternative approach. Consider the observed covariance between populations $i$ and $j$, ${\bf \hat W}_{ij}$. If we had a large number of independent genomic regions and estimated this quantity separately in each independent region, the sampling distribution would be approximately normal with mean ${\bf \hat W}_{ij}$ (by appeal to the central limit theorem). We thus model ${\bf \hat W}_{ij}$ as:

\begin{equation}
{\bf \hat W}_{ij} \sim N({\bf W}_{ij}, \sigma_{ij}^2)
\end{equation}

\noindent where $\sigma_{ij}$ is the standard error in the estimation of ${\bf \hat W}_{ij}$. Because the allele frequencies at nearby SNPs are correlated (i.e., there is linkage disequilibrium), a naive estimate of $\sigma_{ij}$ that treated each SNP as independent would be too small. We instead take a resampling approach to estimate $\sigma_{ij}$. We split the genome into $p$ blocks, such that there are $K$ SNPs per block (with $K$ chosen so that the block sizes are larger than blocks of linkage disequilibrium) \citep{Keinan:2007kx}. (If $K$ does not divide evenly into $n$, we discard the remaining SNPs.) We then calculate $\bf \hat W$ separately in each block. Let  ${\bf \hat W}_{ijk}$ be the sample covariance between two populations $i$ and $j$ in block $k$. Now,
\begin{equation}
 {\bf \hat W}_{ij} = \frac{ \sum_{k=1}^{p} {\bf \hat W}_{ijk}}{ p }
 \end{equation}
 \noindent and
 \begin{equation}
\hat \sigma_{ij} = \sqrt{ \frac{  \sum_{k=1}^{p}  (  {\bf \hat W}_{ijk}-  {\bf \hat W}_{ij})^2}{p(p -1) }}.
\end{equation}

\noindent If we take each pair of populations in turn, we can write down a composite likelihood for $\bf  \hat W$:

\begin{equation} \label{likelihood}
L( {\bf \hat W} | {\bf W}) = \prod \limits_{i=1}^m \prod \limits_{j= i}^m N({\bf \hat W}_{ij} | {\bf W}_{ij}, \hat \sigma_{ij}^2)
\end{equation} 

\noindent where $N({\bf \hat W}_{ij} | {\bf W}_{ij}, \sigma_{ij}^2)$ is a Gaussian density with mean ${\bf W}_{ij}$ and variance $\sigma_{ij}^2$ evaluated at ${\bf \hat W}_{ij}$.

Finally, we wanted to define measures for how well the model fits the data. First, we define the matrix of residuals in this model, $\bf R$. These quantities are useful for visualization and fitting:
\begin{equation}
{\bf R} = {\bf \hat W} - {\bf W}.
\end{equation}

\noindent Positive residuals indicate pairs of populations where the model underestimates the observed covariance, and thus populations where the fit might be improved by adding additional edges. Negative residuals indicate pairs of populations where the model overestimates the observed covariance; these are a necessary outcome of having positive residuals, but can also sometimes be interpreted as populations that are forced too close together due to unmodeled migration elsewhere in the graph. These residuals can be used to define a measure of the fraction of the variance in $\bf \hat W$ that is explained by $\bf W$. Let us call this fraction $f$:
\begin{equation}
f = 1- \frac{ \sum_{i = 1}^m \sum_{j = i+1}^m ({\bf R}_{ij}  - \overline {\bf R})^2}{ \sum_{i = 1}^m \sum_{j = i+1}^m ( {\bf \hat W}_{ij} - {\overline{ \bf \hat W }})^2 } 
\end{equation} 
\noindent where $\overline {\bf R} = \frac{\sum_{i = 1}^m \sum_{j = i+1}^m {\bf R}_{ij}}{m (m-1)/2}$ and $\overline {\bf \hat  W} = \frac{\sum_{i = 1}^m \sum_{j = i+1}^m {\bf \hat W}_{ij}}{m (m-1)/2}$. This quantity approximates the fraction of the variance in relatedness between populations that is accounted for by the model.  

\paragraph{Estimation.} We implemented an algorithm, called \emph{TreeMix}, that searches for the graph that maximizes the composite likelihood in Equation \ref{likelihood}. A search that enumerates all graphs is infeasible unless $m$ is very small, so to simplify the search we make the assumption that the history of the sampled populations is approximately tree-like. We thus start by searching for the maximum likelihood tree, taking an algorithmic approach similar to Felsenstein \citep{Felsenstein_1981}.

After building the tree, we fix the position of the root. (In the tree model the position of the root is not identifiable, as the evolution of allele frequencies along the tree is reversible under the Gaussian model when drift is assumed to be small. In a graph model, though the position of the root is partially identifiable, in all applications we assume that the position of the root is fixed using prior information about known outgroups). We then calculate the residual covariance matrix, $\bf R$, and add migration edges in a directed matter. First, we find the $M$ pairs of populations with the maximum residuals. We then attempt adding a migration edge between populations in the vicinity of each of the $M$ population pairs. For each attempted graph (or tree) topology, we optimize the branch lengths and migration edge weights (Methods). 

After finding the single migration edge that most increases the likelihood, we attempt a series of local changes to the graph structure (Methods). We then iterate over this procedure to add additional migration edges. In principle, migration edges could be added until they are no longer statistically significant (see the following paragraph). In our experience in practice, however, we prefer to stop adding migration events well before this point so that the resulting graph remains interpretable.  

\paragraph{Significance testing.} After building the maximum likelihood graph, we would like to quantify our uncertainty in the resulting graph structure. In particular, we would like to quantify our confidence in individual migration events. However, because the likelihood in Equation \ref{likelihood} is a composite likelihood, it cannot be used directly for formal tests for significance. Instead, we take a resampling approach to test the support for individual migration edges. 

Consider a given migration edge, with corresponding weight $w$. We wish to calculate a p-value for this weight (under the null hypothesis that $w = 0$, and for a fixed graph structure). To do this, we use the Wald statistic $\frac{\hat w}{se(\hat w)}$, where $se(\hat w)$ is the standard error in the estimate of the weight, which is distributed $N(0,1)$ under the null. To obtain the standard error, recall that we have split the genome into $p$ independent blocks. We use the jackknife estimates of both $\hat w$ and the standard error in $\hat w$ (where we jackknife over blocks). Let $i$ index blocks, and $w_{\cdot i}$ be the estimated weight computed using all blocks \emph{except} $i$. Then:
\begin{equation}
\hat w = \frac{ \sum_{i = 1}^{p} \hat w_{\cdot i}}{p}
\end{equation}
\begin{equation}
se(\hat w) = \sqrt{\bigg(\frac{p- 1}{p}\bigg)  \sum_{i = 1}^{p} (\hat w_{\cdot i} - \hat w)^2}
\end{equation}
\noindent This allows us to calculate a p-value for the migration edge. There are a number of complications to the interpretation of this p-value. First, there is the issue of multiple testing--there are at least $2m -2$ edges in the graph (recall that $m$ is the number of populations), and thus around $4m^2$ possible migration events. More importantly, the p-value is generated under a heavily parameterized model: we are comparing a fixed graph structure with a migration event to that same graph without the migration event. A ``significant" p-value simply indicates that the hypothesized migration event significantly improves the fit to the data;  this does not account for the possibility of errors in the graph structure, or indicate that the particular migration event tested is the correct one (rather than a migration event between a different pair of populations). For this reason, we treat the precise p-value generated by this procedure with caution, and use additional, less-parameterized methods like three- and four-population tests \citep{Reich:2009fk} to test the robustness of the inference. 

\subsection*{Simulations}
\begin{figure}
\begin{center}
\includegraphics[scale = 1]{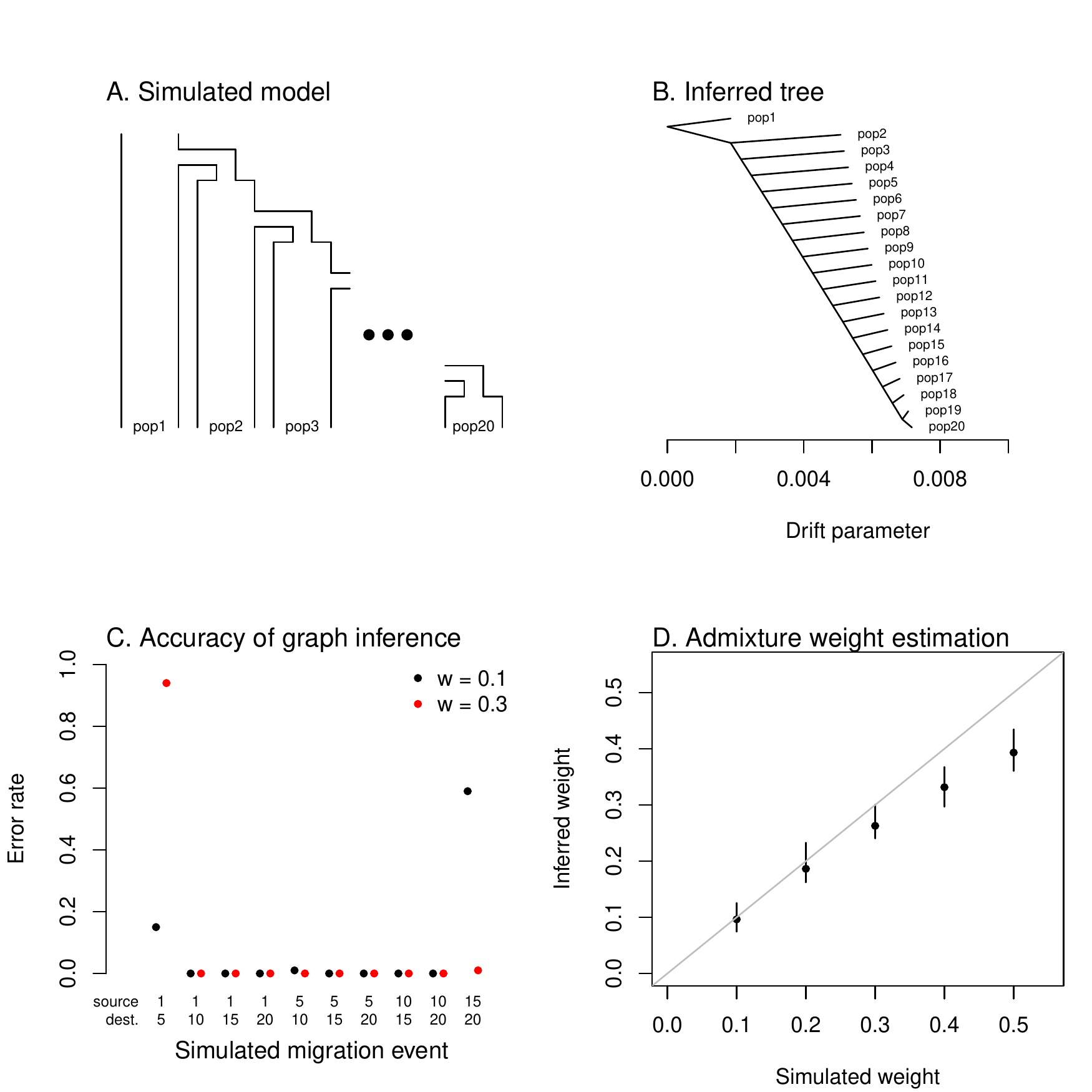}

\caption{: \textbf{Performance on simulated data. A.} The basic outline of the demographic model used. \textbf{B. Trees inferred by \emph{TreeMix}.} We simulated 100 independent data sets, under the demographic model in \textbf{A.}, and inferred the tree. All simulations gave the same topology; plotted are the mean branch lengths. \textbf{C. Performance in the presence of migration.} We added migration events to the tree in \textbf{A.} and inferred the structure of the graph. Each point represents the error rate over 100 independent simulations, defined as the fraction of simulations where the inferred graph topology does not perfectly match the simulated topology. On the x-axis we show the populations involved in the simulated migration event; e.g., if the source population is 1 and the destination population is 10, this is a migration event from population 1 to population 10, as labeled in \textbf{A.} \textbf{D. Admixture weight estimation.} We simulated admixture events with different weights from population 1 to population 10, and inferred the weight. Each point is the mean across 100 simulations, and the bar represents the range.}\label{fig_sim}

\end{center}
\end{figure}

We tested the performance of the \emph{TreeMix} method in simulations. We generated coalescent simulations from several histories; the basic structure was a set of 20 populations produced by a serial bottleneck model like that used by DeGiorgio et al. \citep{DeGiorgio:2009ve} to model human history (Figure \ref{fig_sim}A). The parameters of the simulations were chosen to be reasonable for non-African human populations; we used an effective population size of 10,000, and a history where all 20 populations share a common ancestor 2000 generations in the past. Each individual simulation involved 400 regions of approximately 500kb each, and thus recapitulated many aspects of real data, including hundreds of thousands of loci and the presence of linkage disequilibrium. 

\paragraph{Tree simulations.}
First, we tested the performance of the algorithm on truly tree-like data. We generated 100 independent simulations of 20 chromosomes from each population using the above demographic model without migration, and inferred population trees. The inferred trees perfectly matched the simulated model in all cases (Figure \ref{fig_sim}B, Supplementary Figure 2), and the fitted tree model accounted for an average of 99.8\% of the variance in $\bf \hat W$. To test the effect of SNP ascertainment, we then inferred trees using only SNPs that were polymorphic in one of the populations (either population 1 or population 20); this ascertainment scheme did not decrease accuracy of the inferred topology, though it did alter the inferred branch lengths (Supplementary Figure 3). 

We used these simulations without migration to test the calibration of our p-values for migration events. For each simulation, after building the maximum likelihood tree, we introduced a migration event between two random populations and tested it for significance. As expected if the p-values are properly calibrated, their distribution is approximately uniform (Supplementary Figure 4). 

Finally, we performed tree simulations in a situation where fixed differences and new mutations (rather than shared polymorphisms inherited from a common ancestor) were common between the populations; in this context the population genetic interpretation of the model breaks down. We performed simulations where all the true branch lengths were 50 times longer than in the original model, corresponding to a history where the 20 populations share a common ancestor approximately 100,000 generations in the past. Again, we see no errors in the topology of the inferred trees (Supplementary Figures 5, 6). In this situation, the covariances between closely-related populations tend to be slightly underestimated; in more extreme situations this could lead to spurious inferences of migration (Supplementary Figures 5, 6). However, overall, these simulations suggest that the model will still be useful even in situations where the population genetic interpretation is not strictly correct.

\paragraph{Simulations with migration.} We then introduced migration events into our simulations. We generated simulations under the same model described above; however, we now simulated an admixture event approximately 100 generations before the present where one population receives a fraction of its ancestry (either 10\% or 30\%) from one of the other populations. We tried ten different pairwise combinations of populations, and generated 100 simulations for each pair. For each simulation, we ran \emph{TreeMix} and allowed it to infer a migration event. We then judged the error rate of the algorithm as the fraction of times the inferred topology of the graph was not exactly correct (this is a conservative estimate of the error rate, in that inferred graph topologies that are very close to the truth are counted as errors). In general, \emph{TreeMix} was able to correctly infer the graph structure in these simulations (Figure \ref{fig_sim}C). However, accuracy dropped considerably when migration was between closely related populations without outgroups present in the data (these are populations 1 and 20 in the model; Figure \ref{fig_sim}C). The major types of errors produced in the simulations were incorrectly inferred directions of migration arrows and inference of admixture in populations related to the truly admixed population (Supplementary Material, Supplementary Figure 7). 

We next asked whether the mixture ``weights" inferred in the model can be interpreted as admixture proportions. To do this, we simulated admixture events of varying proportion between the first and tenth population in the serial bottleneck model described above, set the graph to the true topology, and estimated the mixture weight. The weights are indeed correlated with the true ancestry fraction, but underestimate relatively high admixture proportions in these simulations (Figure \ref{fig_sim}D). The precise bias in the estimation of this parameter will depend, in real data, on largely unknowable parameters (Supplementary Material).  

\subsection*{Application to humans}
\begin{figure}
\begin{center}
\includegraphics[scale = 0.7]{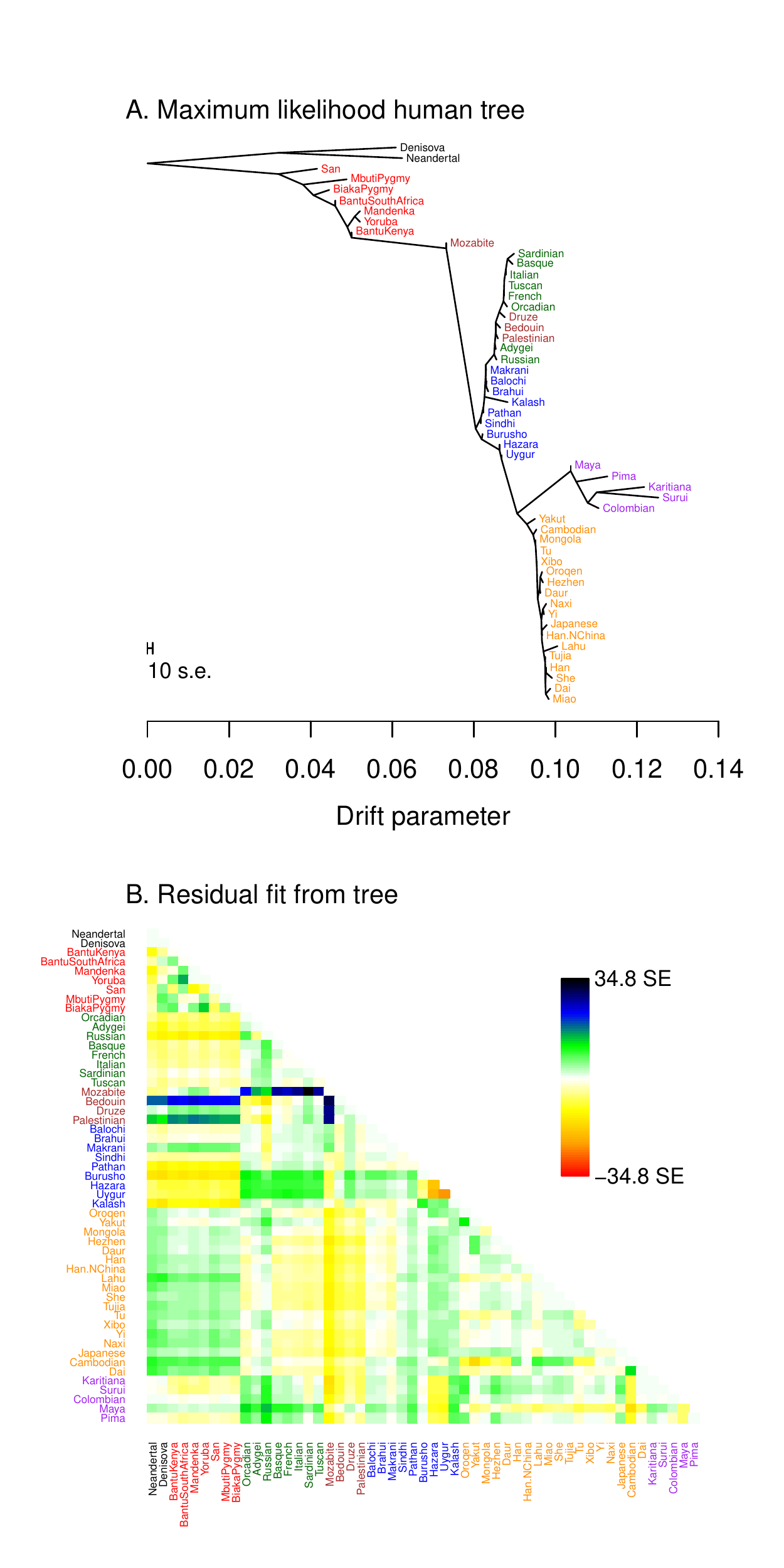}

\caption{\textbf{: Inferred human tree. A. Maximum likelihood tree.} Plotted is the maximum-likelihood tree. Populations are colored according to geographic location (black: archaic humans, red: Africa, brown: Middle East, green: Europe, blue: Central Asia, purple: America, orange: East Asia). The scale bar shows ten times the average standard error of the entries in the sample covariance matrix ($\bf \hat W$). For analysis including Oceania, see Supplementary Figures 11 and 12. \textbf{B. Residual fit.} Plotted is the residual fit from the maximum likelihood tree in \textbf{A.} We divided the residual distance between each pair of populations $i$ and $j$ by the average standard error across all pairs. We then plot in each cell $[i,j]$ this scaled residual. Colors are described in the palette on the right. Residuals above zero represent populations that are more closely related to each other in the data than in the best-fit tree, and thus are candidates for admixture events.}\label{human_tree_fig}

\end{center}
\end{figure}

To test the performance of the \emph{TreeMix} model with real data, we applied it to humans, whose genetic history has been studied extensively \citep{Rosenberg:2002fr, Li:2008rt, Jakobsson:2008fk, Hellenthal:2008zr}. We applied the model to a dataset consisting of about 125,000 SNPs ascertained by low-coverage genome sequencing in a single Yoruban individual and then genotyped in 55 modern and archaic human populations \citep{lu_array}. In all that follows, we excluded the two Oceanian populations because they gave inconsistent results across datasets. We believe this difficulty results from the fact that these populations contain ancestry from multiple sources, making the graph estimation somewhat unstable when they are included (Supplementary Material). We first built the tree of all 53 remaining populations (Figure \ref{human_tree_fig}A). This tree largely recapitulates the known relationships among population groups \citep{Li:2008rt}, and explains 98.8\% of the variance in relatedness between populations (though this is high, it is less than the 99.8\% observed in the simulations of a true tree model). We examined the residuals of the model's fit to identify aspects of ancestry not captured by the tree (Figure \ref{human_tree_fig}B). A number of known admixed populations stand out: in particular, these include the Mozabite and Middle Eastern populations. 

\begin{figure}
\begin{center}
\includegraphics[scale = 1]{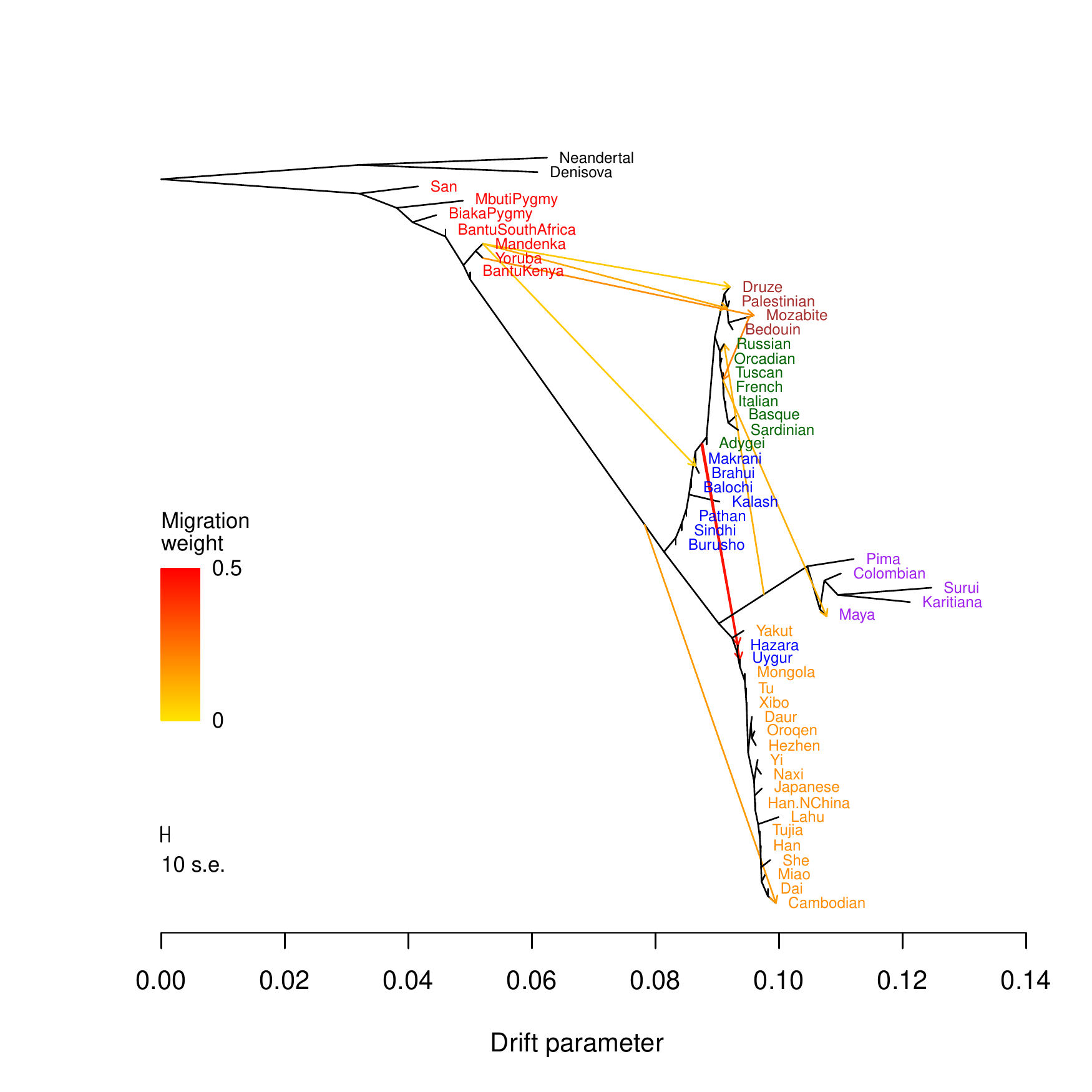}

\caption{\textbf{: Inferred human tree with mixture events.} Plotted is the structure of the graph inferred by \emph{TreeMix} for human populations, allowing ten migration events. Migration arrows are colored according to their weight. Horizontal branch lengths are proportional to the amount of genetic drift that has occurred on the branch. The scale bar shows ten times the average standard error of the entries in the sample covariance matrix ($\bf \hat W$). The residual fit from this graph is shown in Supplementary Figure 9. Admixture from Neandertals to non-African populations is only apparent when considering subsets of the data (see Discussion and Supplementary Figure 15).}\label{human_graph_fig}

\end{center}
\end{figure}

We then sequentially added migration events to the tree. In Figure \ref{human_graph_fig}, we show the inferred graph with ten migration edges; p-values for all reported migration edges are less than $1\times 10^{-30}$ (we show the graph with the maximum likelihood over several independent runs of \emph{TreeMix} with random orders of input populations). This graph model explains 99.8\% of the variance in relatedness between populations. As expected from examination of Figure \ref{human_tree_fig}B, the migration events recapitulate many known events in human history.  We infer that the Mozabite are the result of admixture between an African and a Middle Eastern population (with about 33\% of their ancestry from Africa), and that Middle Eastern populations also have African ancestry  (Palestinians and Bedouins: $w = 13\%$ from Africa; Druze: $w = 6\%$). This is consistent with previously reported admixture proportions from these populations \citep{Moorjani:2011ys, Price:2009vn}. Additionally, we identify the known European ancestry in the Maya ($w = 12\%$) \citep{Rosenberg:2002fr}, and infer that the Uyghur and Hazara populations are the result of admixture between west Eurasian and East Asian populations ($w = 46\%$ and $47\%$ from west Eurasia, respectively) \citep{Xu:2008ys, Rosenberg:2002fr, Lawson:2012fk}.  

Several additional migration events in the human data have not been previously examined in detail, but are consistent with previous clustering analysis of these populations \citep{Rosenberg:2002fr,Lawson:2012fk,Li:2008rt}. These include migration from Africa to the Makrani and Brahui in Central Asia ($w= 5\%$) and from a population related to East Asians and Native Americans (which we interpret as likely Siberian) to Russia ($w = 11\%$). 

Two inferred edges were unexpected. First, perhaps the most surprising inference is that Cambodians trace about 16\% of their ancestry to a population equally related to both Europeans and other East Asians (while the remaining 84\% of their ancestry is related to other southeast Asians). This is partially consistent with clustering analyses, which indicate shared ancestry between Cambodians and central Asian populations \citep{Li:2008rt}. To confirm that the Cambodians are admixed, we turned to less parameterized models. The predicted admixture event implies that allele frequencies in Cambodia are more similar to those in African populations than would be expected based on their East Asian ancestry. To test this, we used three-population tests \citep{Reich:2009fk}. We tested the trees [African, [Cambodian,Dai]] for evidence of admixture in the Cambodians (Methods). When using any African population, there is strong evidence of admixture (when using Yoruba, $Z = -7.0$ [$p  = 1 \times 10^{-12}$]; when using Mandenka, $ Z = -7.3$ [$p = 1\times 10^{-12}$]; when using San, $Z = -4.8$ [$p = 8\times 10^{-7}$]). We conclude that the Cambodian population is the result of an admixture event involving a southeast Asian population related to the Dai and a Eurasian population only distantly related to those present in these data. 

Finally, we infer an admixture edge from the Middle East (a population related to the Mozabite, a Berber population from northern Africa) to southern European populations ($w = 22\%$). This migration edge is the one edge that is not consistent across independent runs of \emph{TreeMix} on these data (Supplementary Figure 8). In particular, an alternative graph (albeit with lower likelihood) places the Mozabite as an admixture between southern Europe and Africa (rather than the Middle East and Africa), and does not include an edge from the middle East to southern Europe. We thus hesitate to interpret this result, except to note that the relationship between northern African, the Middle East, and southern Europe involves complex patterns of gene flow that merit further investigation \cite{Henn:2012fk, Moorjani:2011ys}. 

To test the robustness of our results to SNP ascertainment, we additionally ran \emph{TreeMix} on the same set of populations using a set of SNPs ascertained in a single French individual. The inferred graph was nearly identical (Supplementary Figure 10). Additionally, as noted above, different random input orders for the populations gave very similar results (Supplementary Figure 8). We conclude from this that the model is able to consistently and accurately infer the major mixture events in the history of a species. This approach is computationally efficient: building the tree took around five minutes on a standard desktop computer (with a processor speed of 3.1 GHz), and adding ten migration events to the tree took about four and a half hours (the major computational cost is in the iterative estimation of migration weights).

\subsection*{Application to dogs}
\begin{figure}
\begin{center}
\includegraphics[scale = 0.75]{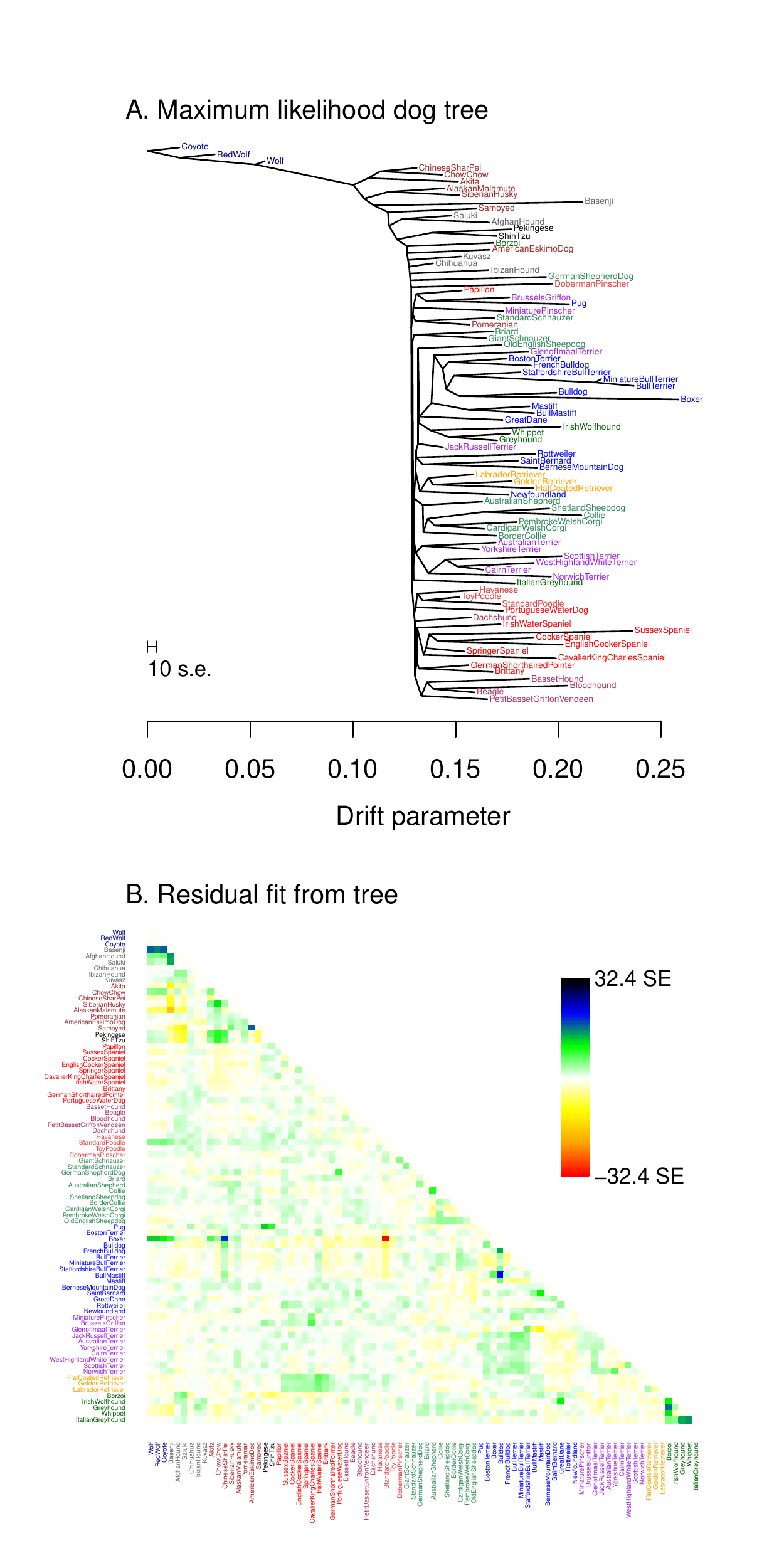}

\caption{\textbf{: Inferred dog tree. A. Maximum likelihood tree.} Populations are colored according to breed type. Dark blue: wild canids, grey: ancient breeds, brown: spitz breeds, black: toy dogs, red: spaniels, maroon: scent hounds,  dark red: working dogs, light green: herding dogs, light blue: mastiff-like dogs, purple: small terriers, orange: retrievers, dark green: sight hounds. The scale bar shows ten times the average standard error of the entries in the sample covariance matrix ($\bf \hat W$).  \textbf{B. Residual fit.} Plotted is the residual fit from the maximum likelihood tree in \textbf{A.} We divided the residual distance between each pair of populations $i$ and $j$ by the average standard error across all pairs. We then plot in each cell $[i,j]$ this scaled residual. Colors are described in the palette on the right. }\label{dog_tree_fig}

\end{center}
\end{figure}

While human populations have been extensively studied, we next applied the model to dogs, a species where considerably less is known about population history. In particular, we applied the model to a dataset consisting of about 60,000 SNPs genotyped in 82 dog breeds or wild canids \citep{Boyko:2010fk}. As for humans, we first inferred the maximum likelihood tree (Figure \ref{dog_tree_fig}A). The differences in history between dogs and humans are striking: there are long terminal branches leading to each dog breed in the inferred tree (Figure \ref{dog_tree_fig}A, recall that the terminal branch lengths account for sample size). This is consistent with the known strong bottlenecks in the establishment of dog breeds \citep{Vonholdt:2010uq}. However, examining the residuals from the model revealed a number of populations that do not fit a strict tree model (Figure \ref{dog_tree_fig}B); indeed, the tree model explained 94.7\% of the variance in relatedness between breeds, somewhat less than between human populations. 

\begin{figure}
\begin{center}
\includegraphics[scale = 1]{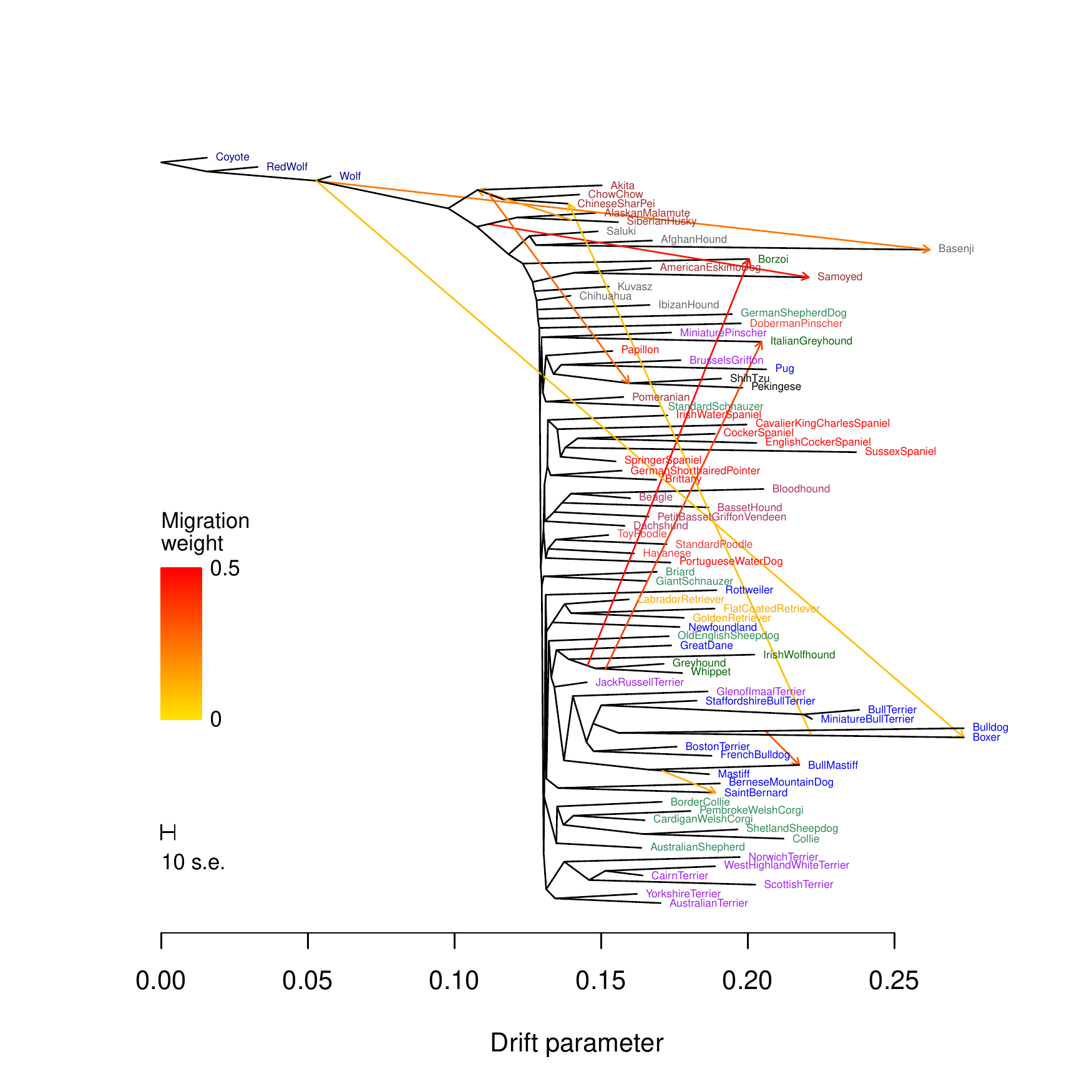}

\caption{\textbf{: Inferred dog graph.} Plotted is the structure of the graph inferred by \emph{TreeMix} for dog populations, allowing ten migration events. Migration arrows are colored according to their weight. The scale bar shows ten times the average standard error of the entries in the sample covariance matrix ($\bf \hat W$). See the main text for discussion. The residual fit from this graph is presented in Supplementary Figure 13.}\label{dog_graph_fig}

\end{center}
\end{figure}

We sequentially added migration events to the tree in Figure \ref{dog_tree_fig}A.  In Figure \ref{dog_graph_fig}, we show the inferred graph with ten migration events, which explains 96.8\% of the variance in relatedness between breeds (which suggests that additional events exist in the data). In the following paragraphs, we describe some of these events. 

We infer that the bull mastiff is the result of an admixture event between bulldogs and mastiffs. This is a known event \citep{american2006complete}; we estimate the admixture proportions as 33\% bulldog and 67\% mastiff. We further examined this event using four-population tests for treeness. As expected given the known history, the tree [[boxer,bulldog],[mastiff,bull mastiff]] fails the four-population test ($Z = 3.5$, $p = 0.002$), while replacing the bull mastiff with other related breeds that we do not predict to be involved in the admixture event results in trees that pass this test. For example, the tree [[boxer,bulldog],[mastiff,Boston terrier]] passes the four-population test with $Z = -0.3$. 

The most visually apparent residuals in Figure \ref{dog_tree_fig}B are accounted for in the graph by an admixture event from the grey wolf into the basenji, an ancient African breed of dog ($w = 25\%$). Such a high mixture fraction is consistent with previous clustering analyses of these data \citep{Vonholdt:2010uq, Parker:2004fk}.  We again sought to confirm this signal in a less-parameterized model. We tested the four-population tree [[wolf,ancient breed],[basenji, Afghan hound]] with various ``ancient" dog breeds. We could not find a tree that passed the four-population test (with Akita as the ancient breed, $Z = 11.7, p < 1\times 10^{-30}$; with Alaskan Malamute, $Z = 13.0, p < 1\times 10^{-30}$), confirming the presence of gene flow in these trees. Replacing the basenji with the saluki in these analyses resulted in trees that pass the four-population test (for example, the tree [[wolf, Akita],[Afghan hound, saluki]] passes with $Z = -0.03, p = 0.51$). Though we cannot have complete confidence in the precise migration events, these results are consistent with admixture between gray wolves and the basenji.

Another breed that stands out in this analysis is the boxer (Note that many of the SNPs used in this study were ascertained using a boxer individual, so we may have increased power to identify migration events involving this breed). We infer a significant genetic contribution from wolves to the boxer ($w = 8\%$), and migration between the boxer and the Chinese shar-pei, a distantly-related ancient breed ($w = 8\%$).  To further examine these events, we again turned to four-population tests. To evaluate the wolf mixture, we tested the tree [[wolf, ancient breed],[boxer, bulldog]]. We did not find a tree that passed the four-population test (with Akita as the ancient breed, $Z = 3.1, p = 0.001$; with Afghan Hound, $Z = 3.4, p  = 0.0003$). Replacing the Boxer with the Mastiff in these analyses led to trees that passed the four-population test (for example, with Akita as the ancient breed, $Z = 0.3, p = 0.38$). To evaluate the gene flow from the Boxer to the Chinese shar-pei, we tested the tree [[Chinese shar-pei, Akita],[boxer, bulldog]]; this tree fails the four-population test ($Z = 3.0, p = 0.001$), while the tree [[Chow Chow, Akita],[boxer,bulldog]] passes  ($Z = -0.48, p = 0.3$). 

Previous analyses of these data have noted that the ``toy breeds" of dog cluster together \citep{Vonholdt:2010uq}. We find that the Chinese toy breeds (the Pekingese and the Shi Tzu) result from admixture between a population related to ancient East Asian dog breeds and a modern population related to the Brussels griffon and the pug ($w = 28\%$ from the East Asian breeds). To confirm the presence of gene flow, we tested four-population trees of the form [[Asian toy breed, Akita/Chow Chow],[Pug,mastiff]]. These trees fail, with varying levels of significance, ranging from [[Chow Chow, Shi Tzu],[Pug, mastiff]] ($Z =  -2.7, p = 0.003$) to [[Akita, Pekingese],[Pug, mastiff]] ($Z = -4.7, p = 1\times 10^{-6}$). 

Finally, we noticed that two of the sighthounds (the Borzoi and the Italian greyhound) do not cluster with the other sight hounds in the tree, namely greyhound, whippet and Irish wolfhound (Figure \ref{dog_tree_fig}A); however, they do show evidence of having sighthound admixture in the graph (Figure \ref{dog_graph_fig}). These are the borzoi (which appear to be admixed between an ancient or spitz-breed dog, with 47\% ancestry from the sighthounds) and the Italian Greyhound (which appears to be admixed with a toy breed, with 34\% ancestry from the sighthounds). This is consistent with the known phenotypic characteristics of these dogs; the borzoi is considered a saluki-like breed, and the Italian greyhound is phenotypically a small version of a greyhound \citep{american2006complete}. 

Overall, we conclude that there has been considerable gene flow between dog breeds over the course of domestication; there are many additional migration events that merit further examination (Figure \ref{dog_graph_fig}, Supplementary Material). 

\section*{Discussion}
In this paper, we have developed a unified model for inferring patterns of population splits and mixtures from genome-wide allele frequency data. We have shown that this model is accurate in simulations, largely recapitulates the known relationships between well-studied human populations, and is able to identify new relationships between populations in both humans and dogs. 

The \emph{TreeMix} model can be thought of as a complement to methods for the identification of population structure \citep{Pritchard:2000zr, Patterson:2006ve, Lawson:2012fk}. These latter methods are powerful tools for clustering together individuals into relatively homogenous populations (and to identify individuals that are genetic outliers in their population)\citep{Pritchard:2000zr, Patterson:2006ve, Lawson:2012fk}. However, once population structure in a species has been identified, these methods are not well-suited for describing \emph{how} it arose, and are only indirectly informative about the historical relationships between different populations. The model developed in this paper is designed to more directly address these historical questions. 

\paragraph{Modeling assumptions.} There are a number of assumptions, both implicit and explicit, in the interpretation of the \emph{TreeMix} model. First, we have motivated the model in terms of inferring the historical splits and mixtures of populations. However, a given covariance structure of allele frequencies between populations can be a consequence of either a non-equilibrium demography (population splits and mixtures) or an equilibrium demography (populations at long-term stasis with a fixed migration structure) \citep{Felsenstein:1982vn}. For the species analyzed in this paper, population equilibrium over the entire species range is not a tenable hypothesis; however, some subsets of populations may be at equilibrium, and there may be species where this alternative historical interpretation of the model is plausible.

We have also modeled migration between populations as occurring at single, instantaneous time points. This is, of course, a dramatic simplification of the migration process. This model will work best when gene flow between populations is restricted to a relatively short time period. Situations of continuous migration violate this assumption and lead to unclear results (Supplementary Figure 14). The relevance of this assumption will depend on the species and the populations considered. In humans, the relevance of continuous versus discrete mixture events is an open question--some aspects of genetic variation appear compatible with continuous migration \citep{Novembre:2008fk}, while other aspects do not \citep{Reich:2009fk}. Indeed, both sorts of models are likely relevant at different time scales \citep{Novembre:2011uq}.  

We have also rely on the implicit assumption that the history of the species being analyzed is largely tree-like. We have made this assumption to simplify the search for the maximum likelihood graph; additionally, we speculate that in graphs with complex structure, there will be many graphs that lead to identical covariance matrices, and thus several different histories will be compatible with the data. That said, improvements to the search algorithm could allow the assumption of approximate treeness to be somewhat relaxed. Currently, if the number of admixed populations is large relative to the number of unadmixed populations, this assumption breaks down. For example, in the human data, note that we see no evidence of the documented gene flow from Neandertals to all non-African populations \citep{Green:2010vn} (Figure \ref{human_tree_fig}B). The reason for this is that the large number of populations with admixture can be accommodated in the tree by allowing the branch from Neandertals to Africans to be slightly underestimated (additionally, by using SNPs ascertained in Africa, we have selected against sites that are informative about Neandertal ancestry). If only a single non-African population is included in the analysis, the relationship between Neandertals and the non-African population is clearer (Supplementary Figure 15).

\paragraph{Conclusions.}
A number of extensions to the sort of model described here are of potential interest. First, the historical relationships between populations could be useful as null demographic models for the detection of natural selection \citep{Coop:2010ly, Bhatia:2011fk, Bonhomme:2010uq}. Second, in a given individual, the best-fit graph relating the individual to other populations may change along a chromosome; this sort of information could be of use in local ancestry inference. Finally, we have not used the information about demographic history present in linkage disequilibrium; approaches that explicitly use this information may provide additional power to detect migration events and estimate their timing, at an additional computational cost \citep{Hellenthal:2008zr, Myers:2011z, Lawson:2012fk}. 

\section*{Methods}

\paragraph{Graph estimation.}
As described in the Results, we developed an algorithm called \emph{TreeMix} that uses the composite likelihood in Equation \ref{likelihood} to search for the maximum likelihood graph. Estimation involves two major steps. First, for a given graph topology, we need to find the maximum likelihood branch lengths and migration weights. Second, we need to search the space of possible graphs. First consider a given graph topology. We iterate between optimizing the branch lengths and weights. If the edge weights are known, the observed entries of the covariance matrix can be written down as an overdetermined system of linear equations (as in Equations 13-15). We solve this system by non-negative least squares \citep{citeulike:5492390}. Though the least squares solution is the maximum composite likelihood solution in the case where all entries of the covariance matrix have equal variance, it is not strictly the maximum likelihood solution in cases with unequal variances. The algorithm could be extended to unequal variances using a weighted least squares approach, but we have not implemented this. We then do a golden section search for the optimal weight (between zero and one) on each migration edge \citep{Press:1992:NRC:148286}. At each step in the golden section search, we update the branch lengths. We optimize the weight of each migration edge in turn, and iterate over migration edges until convergence. 

To search the space of possible graphs, we take a hill-climbing approach. We start by finding a local optimum tree, taking an algorithmic approach similar to Felsenstein \citep{Felsenstein_1981}. We randomly select three populations, optimize the branch lengths for all three possible trees, and choose the best (in terms of the composite likelihood) tree. Then, we add the remaining populations one by one in a random order. To add a population, we try attaching it to all branches of the current tree, optimizing the branch lengths for each one as described above, and find the most likely spot. We then perform a round of local rearrangements (i.e., nearest-neighbor interchanges \citep{citeulike:584535}) around each internal node, keeping the resulting tree only if it increases the likelihood.

After adding all populations, we calculate the residual covariance matrix, $R$. We then add migration edges in a directed matter. First, we find the $M$ pairs of populations with the maximum residuals (these are the pairs of populations with the worst fit under the model). In the results reported, $M = 4$. We define a ``neighborhood" around each population of a pair as the tips within a distance of $E$ edges of the focal population. In applications above, we use $E = 3$. This defines a set of pairs of populations that either have a poor fit, or are located in the graph near populations with a poor fit. We take each of these pairs in turn. For each pair, we identify the set of nodes in the path from each member of the pair to the root of the graph. This gives us two sets of nodes. We take all pairwise combinations of nodes in each set, and look at residuals between the populations that are the descendants of each node. If all of the residuals are positive, we add a migration edge between the two nodes and estimate its maximum likelihood weight. We then keep only the single edge that most increases the likelihood of the graph. After adding a migration edge, we attempt nearest-neighbor interchanges at the source and destination of the migration event, attempt changing the source and destination of all migration events, and attempt changing the direction of all migration arrows. Once we have reached the local maximum by this method, we attempt nearest-neighbor interchanges at all internal nodes. We iterate over this procedure for a predetermined number of migration edges. We then test the migration edges for significance as described. 

The \emph{TreeMix} source code is available at \url{http://treemix.googlecode.com}.

\paragraph{Three- and four-population tests of treeness.} We implemented three- and four-population tests as described in Reich et al. \citep{Reich:2009fk}. For the relationship between the $f-$statistics and the covariance model underlying \emph{TreeMix}, see the Supplementary Material. For the three-population test, we estimated $f_3$ as in Reich et al. \citep{Reich:2009fk}, and tested whether is it less than zero. We report the Z-score for this test. To obtain a standard error on the estimate of $f_3$, we used a block jackknife similar to Reich et al. \citep{Reich:2009fk}. However, Reich et al. \citep{Reich:2009fk} split the genome into blocks based on distance (with variable numbers of SNPs per block); we split the genome into blocks of $K$ SNPs (and thus the blocks will be of variable size). 

For the four-population test for treeness, we calculate the $f_4$ statistic as in Reich et al. \citep{Reich:2009fk}, and test whether it is different than zero. Again, we report a Z-score for this test. Standard errors for the $f_4$ statistic were obtained as for the $f_3$ statistic. 
 
\paragraph{Human data.} The human data we used were downloaded from \url{http://www.cephb.fr/en/hgdp/} on August 16th, 2011 (the data set labeled Harvard HGDP-CEPH genotypes). They consist of several panels of SNPs ascertained from low-coverage genome sequencing of single individual from different populations and then genotyped in the Human Genome Diversity Panel \citep{lu_array}. Additionally, at each site, a single sequencing read from the Denisova and Neandertal genome sequencing projects was sampled and the allele reported. These data have the property that they allow for complete control of the ascertainment strategy, and allow us to test the robustness of inference to different ascertainment schemes. For the main analyses, we used the panel of autosomal SNPs ascertained in a single Yoruban individual; there are 124,115 such sites. For some analyses, we also used the panel of autosomal SNPs ascertained in a single French individual; there are 111,970 such sites. For all analyses with \emph{TreeMix}, we used a window size ($-K$) of 500; this corresponds to a window size of  approximately 10Mb. For all \emph{TreeMix} analyses, we set the Neandertal and Denisova samples as the outgroups. 

Since we have only a single allele from the Neandertal and Denisova populations, we cannot calculate heterozygosity in these populations for unbiased estimation of the covariance matrix (see Supplementary Information). To account for this, we simply chose a relatively low level of heterozygosity and assigned it to both populations. In the Yoruba ascertained SNPs, we used a heterozygosity of 0.13, and for the French ascertained SNPs, we used a heterozygosity of 0.2. In practice, this only affected the lengths of the terminal branches to Neandertal and Denisova; running \emph{TreeMix} with a heterozygosity of zero in both populations resulted in the same graph topologies (not shown). 

\paragraph{Dog data.} Allele counts for the dog breeds and wild canids reported in Boyko et al. \citep{Boyko:2010fk} were downloaded from \url{http://genome-mirror.bscb.cornell.edu/} on July 30, 2011. These data consist of counts of reference and alternate alleles at 61,468 sites in 85 dog breeds and wild canids. We removed the Jackal and Scottish Deerhound for having relatively high amounts of missing data, and the village dogs because it is unclear if they represent a coherent population. We also removed all SNPs on the X chromosome. This left us with 60,615 SNPs in 82 populations. We ran \emph{TreeMix} with a window size ($-K$) of 500. This corresponds to a window size of approximately 20 Mb. For all \emph{TreeMix} analyses, we set the coyote as the outgroup. 

The ascertainment scheme used for SNP discovery in dogs was complicated \citep{Lindblad-Toh:2005kx}. The largest set of SNPs were ascertained by virtue of being different between the boxer and poodle assemblies. This should lead to an overestimation of the distance between the boxer and the poodle in our analysis. Indeed, in Figure \ref{dog_tree_fig}B, a considerable negative residual between the boxer and poodle is visible. Another set of SNPs were ascertained by being heterozygous within a boxer individual, and a third set were ascertained by comparison between a boxer and wild canids. These latter SNPs should lead to an overestimation of the distance between the boxer and the wolf in our analysis (as we see for the poodle); in fact, we infer migration between the boxer and the wolf. This ascertainment issue may have led us to underestimate the amount of gene flow in the comparison. 

\paragraph{Simulations.} All simulations were performed using \emph{ms} \citep{Hudson:2002ys}. The exact commands used are listed in the Supplementary Material. When running \emph{TreeMix} on simulations without ascertainment, we used a window size of 5000 SNPs; for simulations with ascertainment we used windows of 1000 SNPs. Consensus trees were generated using SumTrees v.3.1.0. \citep{Sukumaran:2010fk}

\paragraph{Acknowledgements.} We thank three anonymous reviewers, David Reich, Nick Patterson, Graham Coop, Peter Ralph, Daniel Falush, and Daniel Lawson for helpful comments and suggestions. 
\clearpage
\bibliography{bib}

\end{document}


\title{Supplementary Material for: Inference of population splits and mixtures from genome-wide allele frequency data}
 \author{Joseph K. Pickrell$^{1,3,\dagger}$, Jonathan K. Pritchard$^{1,2,\dagger}$\\ 
 \\
 \small $^1$ Department of Human Genetics and\\
 \small $^2$ Howard Hughes Medical Institute, University of Chicago\\
 \small $^3$ Current address: Department of Genetics, Harvard Medical School\\
 \small $\dagger$ To whom correspondence should be addressed: joseph\_pickrell@hms.harvard.edu, pritch@uchicago.edu
 }

\maketitle

\paragraph{Correcting covariances for finite sample size.} In the main text, we define the variance\- covariance matrix $\bf \hat W$  of allele frequencies between populations without accounting for sampling variance. Here, we show the calculations corrected for sample size. Consider $n$ biallelic loci typed in $m$ populations of diploid individuals, and let the sample size in population $i$ at locus $k$ be $N_{ik}$ (with missing data, the number of individuals can vary across loci).  Let the counts of the two alleles in population $i$ at locus $k$ be $n_{ik}$ and $2N_{ik} - n_{ik}$ (with one allele being arbitrarily defined as the reference in all that follows), the true allele frequency in the population be $X_{ik}$, and the observed allele frequency be $\hat X_{ik} = \frac{n_{ik}}{2N_{ik}}$. We assume the $n_{ik}$ are binomially distributed with parameters $2N_{ik}$ and $X_{ik}$, and are independent for all $i$ and $k$. Recall that the allele frequency in the ancestral population is $x_A$, and that the covariance between populations $i$ and $j$ with respect to the ancestral frequency $x_A$ is $ {\bf V}_{ij}$. We begin by defining ${\bf V}_{ij}$ using the observed allele frequencies at a single SNP $k$:
\begin{align}
{\bf V}_{ij} &= E[(\hat X_{ik} - x_A)(\hat X_{jk}-x_A)]\\
&= E\big[   [ (\hat X_{ik}-X_{ik})+(X_{ik} - x_A )][ (\hat X_{jk}-X_{jk}  )+(X_{jk} - x_A)]\big]\\
&=  E[(X_{ik} -x_A)(X_{jk} - x_A)]+ E[  (\hat X_{ik} - X_{ik})(\hat X_{jk} -  X_{jk})].
\end{align}
The bias in the estimate of ${\bf V}_{ij}$ is thus $E[ (\hat X_{ik} - X_{ik})^2]$ if $i = j$ (i.e., it is the sampling variance in $X_{ik}$) and zero otherwise.  This follows from the fact that the $n_{ik}$ are assumed to be independent across $i$. 

Now consider all $n$ SNPs, and let the mean bias across all SNPs be $B_i$. At a given SNP $k$, the sampling variance in population $j$ is $\hat X_{ik}$ is $\frac{X_{ik}(1-X_{ik})}{2N_{ik}}$ (from the binomial sampling of $x_{ik}$), so the mean bias across SNPs is proportional to $\overline {X_{ik}(1-X_{ik})}$ (i.e., the mean across all SNPs of $X_{ik}(1-X_{ik})$). A natural estimator of $B_i$ is then:
\begin{equation}
B_i = \frac{h_i}{4 N_i}
\end{equation}
where $h_i$ is an unbiased estimate of the heterozygosity in population $i$ averaged over all SNPs \citep{Nei:1978fk}:
\begin{equation}
h_i = \frac{1}{n} \sum_{k = 1}^n \frac{n_{ik} (2N_i - n_{ik})}{ N_i (2N_i-1)}.
\end{equation}
As derived in the main text, the sample covariance of populations $i$ and $j$, ${\bf W}_{ij}$, is:
\begin{equation}
{\bf W}_{ij}  = {\bf V}_{ij} - \frac{1}{m} \sum_{k = 1}^m {\bf V}_{ik} - \frac{1}{m} \sum_{k = 1}^m {\bf V}_{jk} + \frac{1}{m^2} \sum_{k = 1}^m \sum_{k' = 1}^m {\bf V}_{kk'}.
\end{equation}
The bias in the estimate of ${\bf \hat W}_{ij}$ (let us call this $B'_{ij}$) is then:
\begin{equation}
B'_{ij} = I_{[i=j]} B_i - \frac{B_i}{m} - \frac{B_j}{m} + \frac{ \sum_{k = 1}^m B_k}{m^2}
\end{equation}
where $I_{[i=j]}$ is an indicator that evaluates to 1 if $i = j$ and zero otherwise. We can then estimate the unbiased covariance ${\bf \hat W}_{ij}$ as:
\begin{equation}
{\bf \hat W}_{ij} = \frac{ \sum_{k = 1}^n (\hat X_{ik} - \mu_k)( \hat X_{jk} - \mu_k) }{n } - B'_{ij}
\end{equation}
where $\mu_k =\frac{ \sum_{i = 1}^m \hat X_{ik} }{m}$. If there is missing data in either population $i$ or population $j$, we simply ignore the SNP for that pairwise comparison of populations. Since the mean allele frequency across populations is important here, large amounts of missing data (or correlated missingness between populations) could result in skewed covariances. We thus exclude populations with large amounts of missing data. 

\paragraph{Nonidentifiability of the drift parameters in an admixed population.} In the main text, we write down a model for the allele frequencies in an admixed population, and claim that the amount of genetic drift occurring before and after the mixture event are nonidentifiable. Consider the graph in Supplementary Figure \ref{fig_mix}. We can write down the expected variances and covariances involving the admixed population:
\begin{align}
{\bf V}_{12} &= (1-w) c_4 x_A[1-x_A]\\
{\bf V}_{23} &= w c_5 x_A[1-x_A]\\
{\bf V}_{22} &= [c_1 + w^2 (c_2+c_5) + (1-w)^2 (c_3+c4)] x_A[1-x_A]
\end{align} 
\noindent and we are interested in estimating $w$, $c_1$, $c_2$, and $c_3$. It is clear from the above that $c_1$, $c_2$, and $c_3$ do not appear except as a linear combination. Adding additional populations does not add additional information about these parameters, unless they are assumed to result from the same mixture event.  

We choose to set $c_2$ and $c_1$ to zero, and estimate only $c_3$, which can now be thought of as a composite branch length that sums all the three components of genetic drift. A subtle point is that all of this drift is weighted by $(1-w)$. When estimating $w$, then, the true relative contributions of $c_1$, $c_2$, and $c_3$ could lead to a bias in the estimation of $w$. For example, if $c_1$ and/or $c_2$ are large, this could bias the estimation of $(1-w)$ upwards. We believe this is likely the cause of the downward bias in $w$ in the simulations in Figure 2D in the main text.

\paragraph{Graph representation of the \emph{TreeMix} model.}
In the main text, we describe a specification of $\bf V$ (the variance/covariance matrix of allele frequencies, defined with respect to an ancestral population) in terms of a system of linear equations. A useful alternate notation describes $\bf V$ in terms of a graph \citep{koller2009probabilistic}. Let $G$ be a rooted, directed, acyclic graph with a set of nodes $N$ and a set of directed edges $E$. Each edge $e$ has an associated length, $c_e$, and a weight, $w_e$ (between zero and one). A special class of edges, called migration edges, are forced to have length zero. The sum of weights of edges entering a given node is one. There is one node which is the root (a node with only outgoing edges), and each population corresponds to a tip (a node with only incoming edges).

Define $\{P_i\}$ to be the set of all possible paths in $G$ from the root to the tip corresponding to population $i$ (if the graph is a tree, there is only one such path). Each individual path $p$ has a weight, $w(p) = \Pi_{e \in p} w_e$. Now define the overlap between two paths as:

\begin{equation}
O(p_i, p_j) = \sum_{e \in p_i} w(p_i) w(p_j) I[e \in p_j] c_e
\end{equation}

\noindent where $I[e \in p_j]$  is a function that evaluates to one if edge $e$ is in $p_j$, and zero otherwise. We can now write down the expected covariance between populations $i$ and $j$ as:
 
\begin{equation} \label{eq_var}
{\bf V}_{ij} = \sum_{p_i \in \{P_i\}} \sum_{p_j \in \{P_j\}} O(p_i, p_j).
\end{equation}

\noindent In the special case where $G$ is a tree, there is only one path per population and all of the edges have weight one, and so ${\bf V}_{ij}$ reduces to a sum of the lengths of branches shared by the two populations.

\paragraph{Relationship of this model to $f-$ statistics.}
Tests for ``treeness" in three and four-population trees \citep{Reich:2009fk, Keinan:2007kx} have used a framework in which the distances between populations are quantified in terms of ``$f-$statistics" comparing the allele frequencies between the populations. Below, we briefly describe these tests in the notation of our model. Consider the expected $f_3$ statistic calculated between populations $1$, $2$, and $3$, with corresponding allele frequencies $X_1$, $X_2$, and $X_3$. 

\begin{align}
f_3(X_1;  X_2, X_3) &= E[(X_1 - X_2)(X_1-X_3)]\\
&=E\big[[ (X_1 - x_A)-(X_2-x_A)][ (X_1-x_A)-(X_3- x_A)]\big]\\
&= {\bf V}_{11} - {\bf  V}_{12}- {\bf V}_{13} + {\bf V}_{23}.
\end{align}
\noindent Consider the situation where populations $1$ and $3$ form a clade relative to $2$ (i.e., population $2$ is an outgroup). If population $X_1$ is not admixed, this reduces to:
\begin{equation}
f_3(X_1;  X_2, X_3) = {\bf V}_{11} -  {\bf V}_{13}.
\end{equation}
\noindent This is necessarily greater than zero (since ${\bf V}_{13} <= {\bf V}_{11}$). If $X_1$ is admixed, then ${\bf  V}_{12}$ can be important and the $f_3$ statistic can be negative. A test for a negative $f_3$ statistic is thus a test for admixture in population $X_1$ \citep{Reich:2009fk}.  However, this signal can be weakened by large amounts of drift in $X_1$ (i.e., a large ${\bf V}_{11}$), or mixture between $X_2$ and $X_3$ \citep{Reich:2009fk}.

Similarly, consider the expected $f_4$ statistic computed on the tree [[1,2],[3,4]], where $1$, $2$, $3$, and $4$ are populations, and $X_1$, $X_2$, $X_3$ and $X_4$ are the corresponding allele frequencies:
\begin{align}
f_4(X_1, X_2;  X_3, X_4) &= E[(X_1 - X_2)(X_3-X_4)]\\
&=E\big[[ (X_1 - x_A)-(X_2-x_A)][ (X_3-x_A)-(X_4- x_A)]\big]\\
&=  {\bf V}_{13}-  {\bf V}_{23} - {\bf V}_{14} + {\bf V}_{24} \\
\end{align}
\noindent If the tree is correct (i.e., if populations $1$ and $2$ are a clade relative to populations $3$ and $4$), all of these quantities are zero. A test for a non-zero $f_4$ statistic is thus a test for treeness \citep{Reich:2009fk}.

\paragraph{Simulation commands.} For all simulations, we used \emph{ms} \citep{Hudson:2002ys}. To generate the tree-like data depicted in Figure 2A in the main text, the command is: \\
\\
\texttt{ms 400 400 -t 200 -r 200 500000 -I 20 20 20 20 20 20 20 20 20 20 20 20 20 20 20 20 20 20 20 20 20 -en 0.00270 20 0.025 -ej 0.00275 20 19 -en 0.00545 19 0.025 -ej 0.00550 19 18 -en 0.00820 18 0.025 -ej 0.00825 18 17 -en 0.01095 17 0.025 -ej 0.011 17 16 -en 0.01370 16 0.025 -ej 0.01375 16 15 -en 0.01645 15 0.025 -ej 0.01650 15 14 -en 0.01920 14 0.025 -ej 0.01925 14 13 -en 0.02195 13 0.025 -ej 0.02200 13 12 -en 0.02470 12 0.025 -ej 0.02475 12 11 -en 0.02745 11 0.025 -ej 0.02750 11 10 -en 0.03020 10 0.025 -ej 0.03025 10 9 -en 0.03295 9 0.025 -ej 0.03300 9 8 -en 0.03570 8 0.025 -ej 0.03575 8 7 -en 0.03845 7 0.025 -ej 0.03850 7 6 -en 0.04120 6 0.025 -ej 0.04125 6 5 -en 0.04395 5 0.025 -ej 0.04400 5 4 -en 0.04670 4 0.025 -ej 0.04675 4 3 -en 0.04945 3 0.025 -ej 0.04950 3 2 -en 0.05220 2 0.025 -ej 0.05225 2 1}
\\
\\
To create trees with considerably longer branch lengths, we multiplied all branch lengths by 50:
\\
\texttt{ms 400 400 -t 200 -r 200 500000 -I 20 20 20 20 20 20 20 20 20 20 20 20 20 20 20 20 20 20 20 20 20 -en 0.135 20 0.025 -ej 0.1375 20 19 -en 0.2725 19 0.025 -ej 0.275 19 18 -en 0.41 18 0.025 -ej 0.4125 18 17 -en 0.5475 17 0.025 -ej 0.55 17 16 -en 0.685 16 0.025 -ej 0.6875 16 15 -en 0.8225 15 0.025 -ej 0.825 15 14 -en 0.96 14 0.025 -ej 0.9625 14 13 -en 1.0975 13 0.025 -ej 1.1 13 12 -en 1.235 12 0.025 -ej 1.2375 12 11 -en 1.3725 11 0.025 -ej 1.375 11 10 -en 1.51 10 0.025 -ej 1.5125 10 9 -en 1.6475 9 0.025 -ej 1.65 9 8 -en 1.785 8 0.025 -ej 1.7875 8 7 -en 1.9225 7 0.025 -ej 1.925 7 6 -en 2.06 6 0.025 -ej 2.0625 6 5 -en 2.1975 5 0.025 -ej 2.2 5 4 -en 2.335 4 0.025 -ej 2.3375 4 3 -en 2.4725 3 0.025 -ej 2.475 3 2 -en 2.61 2 0.025 -ej 2.6125 2 1}
\\
\\
For simulations with migration, we added a migration event approximately 100 generations before the present. For example, migration from population 1 to population 10:
\\
\\
\texttt{ ms 400 400 -t 200 -r 200 500000 -I 20 20 20 20 20 20 20 20 20 20 20 20 20 20 20 20 20 20 20 20 20 -em 0.002675 10 1 4000 -en 0.00270 20 0.025 -em 0.00270 10 1 0 -ej 0.00275 20 19 -en 0.00545 19 0.025 -ej 0.00550 19 18 -en 0.00820 18 0.025 -ej 0.00825 18 17 -en 0.01095 17 0.025 -ej 0.011 17 16 -en 0.01370 16 0.025 -ej 0.01375 16 15 -en 0.01645 15 0.025 -ej 0.01650 15 14 -en 0.01920 14 0.025 -ej 0.01925 14 13 -en 0.02195 13 0.025 -ej 0.02200 13 12 -en 0.02470 12 0.025 -ej 0.02475 12 11 -en 0.02745 11 0.025 -ej 0.02750 11 10 -en 0.03020 10 0.025 -ej 0.03025 10 9 -en 0.03295 9 0.025 -ej 0.03300 9 8 -en 0.03570 8 0.025 -ej 0.03575 8 7 -en 0.03845 7 0.025 -ej 0.03850 7 6 -en 0.04120 6 0.025 -ej 0.04125 6 5 -en 0.04395 5 0.025 -ej 0.04400 5 4 -en 0.04670 4 0.025 -ej 0.04675 4 3 -en 0.04945 3 0.025 -ej 0.04950 3 2 -en 0.05220 2 0.025 -ej 0.05225 2 1}
\\
\\
For simulations of populations exchanging migrants on a lattice, we used the following command:
\\
\\
\texttt{  ms 200 1500 -t 40 -r 40 100000 -I 10 20 20 20 20 20 20 20 20 20 20 -ma x 0 0 0 0 0 0 0 0 0 0 x 0.1 0 0.1 0.1 0 0 0 0 0 0.1 x 0.1 0.1 0.1 0.1 0 0 0 0 0 0.1 x 0 0.1 0.1 0 0 0 0 0.1 0.1 0 x 0.1 0 0.1 0.1 0 0 0.1 0.1 0.1 0.1 x 0.1 0.1 0.1 0.1 0 0 0.1 0.1 0 0.1 x 0 0.1 0.1 0 0 0 0 0.1 0.1 0 x 0.1 0 0 0 0 0 0.1 0.1 0.1 0.1 x 0.1 0 0 0 0 0 0.1 0.1 0 0.1 x -ej 0.41 10 1 -ej 0.41 9 1 -ej 0.41 8 1 -ej 0.41 7 1 -ej 0.41 6 1 -ej 0.41 5 1 -ej 0.41 4 1 -ej 0.41 3 1 -ej 0.41 2 1}
\\
\\
\paragraph{Discussion of simulation errors.} In Figure 2C in the main text, we showed that \emph{TreeMix} was extremely accurate in most simulation situations. However, there are a few situations in which it performed poorly. Most notably, this was for simulated admixture between population 1 and 5. The errors in these simulations tended to be of the same type (Supplementary Figure \ref{fig_error}). Additionally, in the simulations of migration from population 15 to population 20 with a weight of 10\%, there was also a considerable error rate. However, these errors were not consistent across simulations, and are likely due to the algorithm simply not detecting the admixture event at all. 

\paragraph{Analysis of human data including Oceanian populations.} As described in the main text, in the human HGDP data we used two sets of allele frequencies with different ascertainment schemes--one at SNPs ascertained by sequencing a single Yoruban individual, and one at SNPs ascertained by sequencing a single French individual. We initially ran \emph{TreeMix} on both data sets using all populations to estimate the maximum likelihood trees. The trees estimated using the two ascertainment schemes are nearly identical (Supplementary Figure \ref{fig_woc_tree}). We then used \emph{TreeMix} to identify migration events. The algorithm arrived at quite different conclusions about the Oceanian populations in the two different data sets (recall that these are the exact same individuals, just genotyped at different SNPs) (Supplementary Figure \ref{fig_woc_graphs}). In the Yoruba-ascertained data, the East Asian populations are inferred to be admixed, with the Melanesians as a source population. However, in the French-ascertained data, the Oceanians are inferred to be admixed. When the Oceanian populations are excluded from analysis, the algorithm comes up with nearly the same graph in both datasets (Figure 4 in the main text and Supplementary Figure \ref{fig_fr}). 

It is not immediately clear why there is a discrepancy between these two datasets when looking at Oceanian populations. However, Oceania has a particularly complicated genetic makeup, involving at least four distinct components of ancestry: Denisovan gene flow, Neandertal gene flow, native Oceanian, and gene flow from Austronesian speakers
\citep{Reich:2011uq, Reich:2010uq, Wollstein:2010fk}. These different components of ancestry may be picked up to differing extents by SNPs from the different ascertainment panels, leading to conflicting results.

\paragraph{List of migration events inferred in the human data.} Here we list the ten migration edges inferred in the human data and present in Figure 4 in the main text:

\begin{enumerate}
\item Yoruba $\rightarrow$ Mozabite, $w = 19\%$
\item Mandenka $\rightarrow$ (Palestinian,(Bedouin, Mozabite)), $w = 13\%$
\item Mandenka $\rightarrow$ Druze, $w = 6\%$
\item Mankenka $\rightarrow$ (Brahui, Makrani), $w = 6\%$
\item Ancestral non-African $\rightarrow$ Cambodian, $w = 17\%$
\item (All Europe, Middle East) $\rightarrow$ Hazara, $w = 46\%$
\item (All Europe, Middle East) $\rightarrow$ Uygur, $w = 46\%$
\item All Native Americans $\rightarrow$ Russian, $w = 12\%$
\item  (Tuscan(French(Italian,(Basque,Sardinian) $\rightarrow$ Maya, $w = 12\%$
\item Mozabite $\rightarrow$ (Tuscan(French(Italian,(Basque,Sardinian), $w = 22\%$
\end{enumerate}

\paragraph{List of migration events inferred in the dog data.} Here we list the ten migration edges inferred in the dog data and present in Figure 6 in the main text:

\begin{enumerate}
\item (Greyhound, Whippet) $\rightarrow$ Borzoi, $w = 47\%$
\item (AlaskanMalamute,SiberianHusky) $\rightarrow$ Samoyed, $w = 47\%$
\item Wolf $\rightarrow$ Basenji, $w = 25\%$
\item (ChowChow,ChineseSharPei) $\rightarrow$ (Pekingese, Shih Tzu), $w = 28\%$
\item Bulldog $\rightarrow$ Bull mastiff, $w = 31\%$
\item Boxer $\rightarrow$ Chinese shar-pei, $w = 8\%$
\item Siberian Husky $\rightarrow$  (Akita,(ChowChow,ChineseSharPei)) $w = 17\%$
\item Wolf $\rightarrow$ Boxer, $w = 8\%$
\item Mastiff$\rightarrow$ Saint Bernard, $w = 13\%$
\item Whippet $\rightarrow$ Italian Greyhound, $w = 37\%$
\end{enumerate} 
\begin{figure}
\begin{center}
\includegraphics[scale = 0.7]{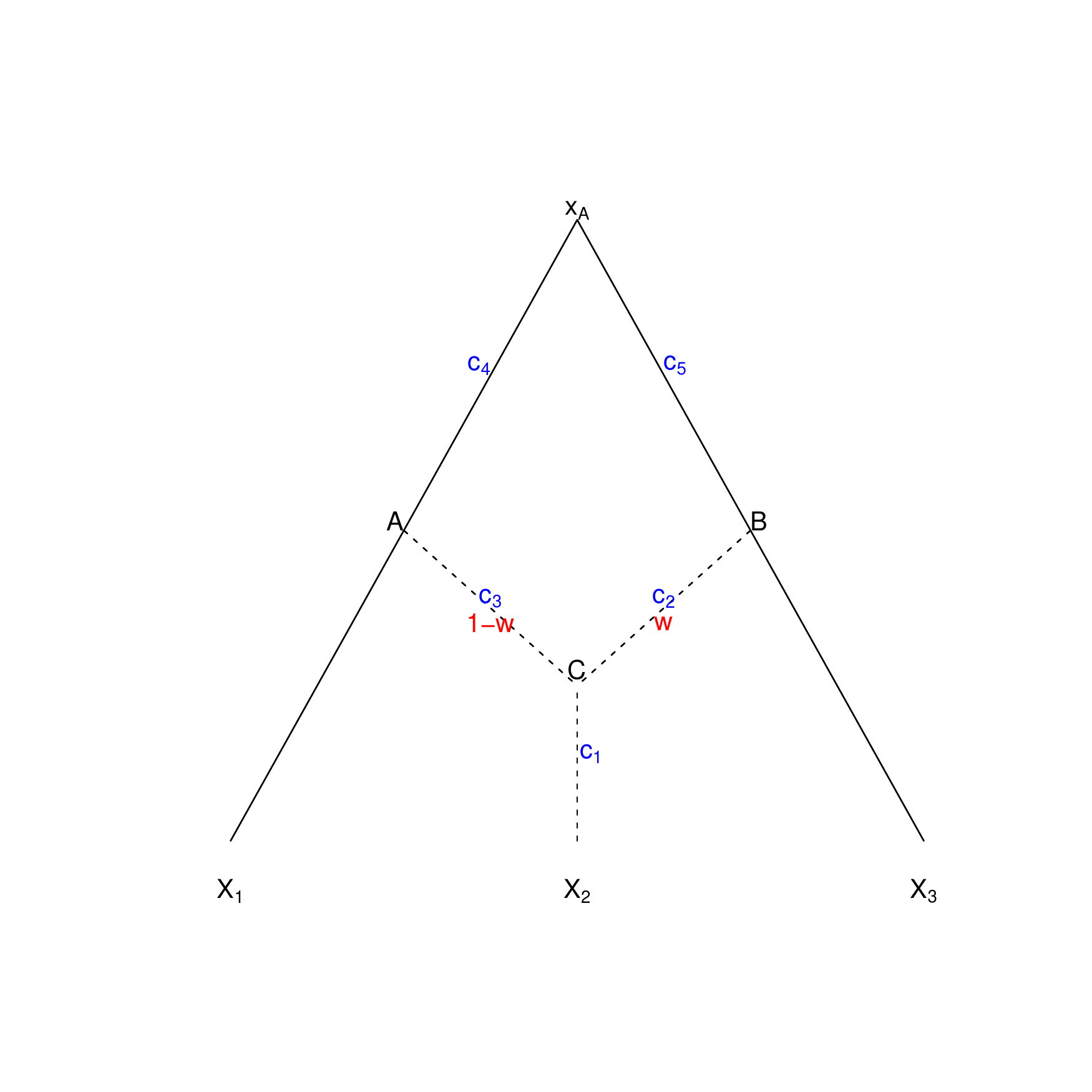}

\caption{\textbf{: A graph with a mixture event.} Capital letters represent nodes, branch length parameters are in blue, and weight parameters are in red.}\label{fig_mix}

\end{center}
\end{figure}

\begin{figure}
\begin{center}
\includegraphics[scale = 1]{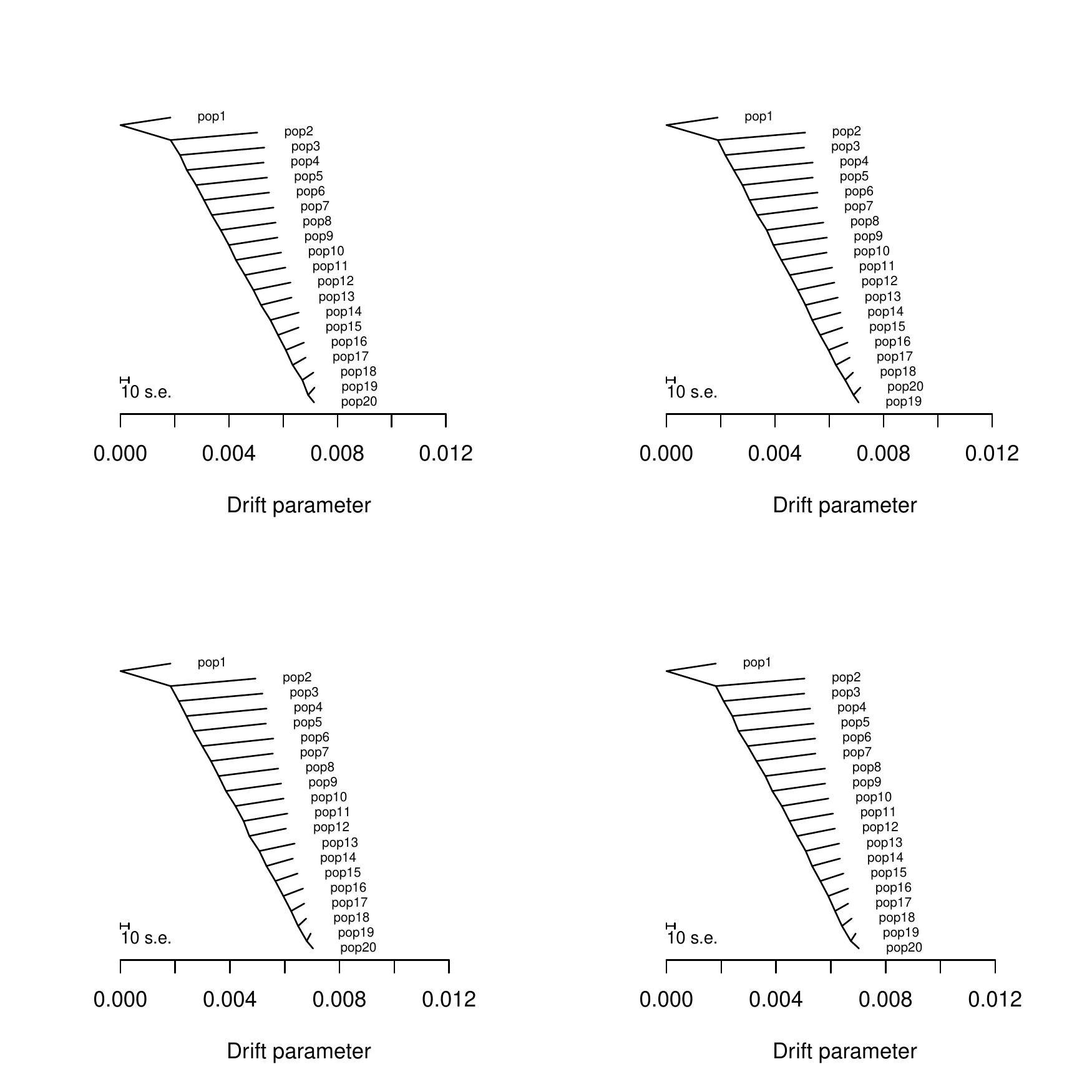}

\caption{\textbf{: Replicates of inferred trees from simulated data.} We generated tree-like data using the topology in Figure 2A in the main text. In figure 2B in the main text, we show the inferred tree with mean branch lengths. In this figure, we show four representative individual trees.}\label{fig_rep_sim}

\end{center}
\end{figure}

\begin{figure}
\begin{center}
\includegraphics[scale = 1]{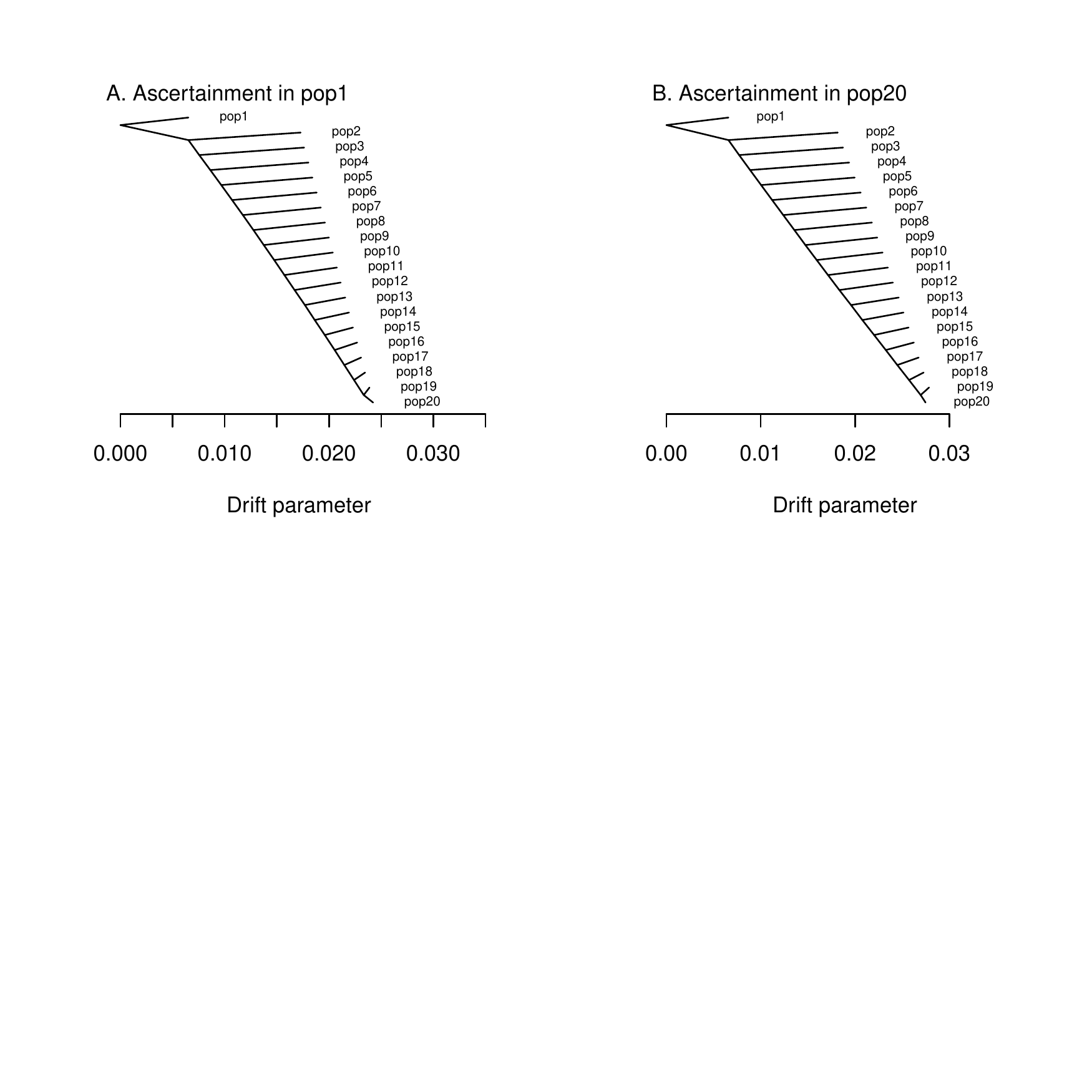}

\caption{\textbf{: Inferred trees on ascertained data.} We generated tree-like data using the topology in Figure 2A in the main text. We then used only the SNPs that were polymorphic in either population 1 (\textbf{A.}) or population 20 (\textbf{B.}) to infer the trees. The correct topology was obtained in all 100 simulations; the branch lengths in each figure are the mean across all simulations.}\label{fig_asc}

\end{center}
\end{figure}

\begin{figure}
\begin{center}
\includegraphics[scale = 0.7]{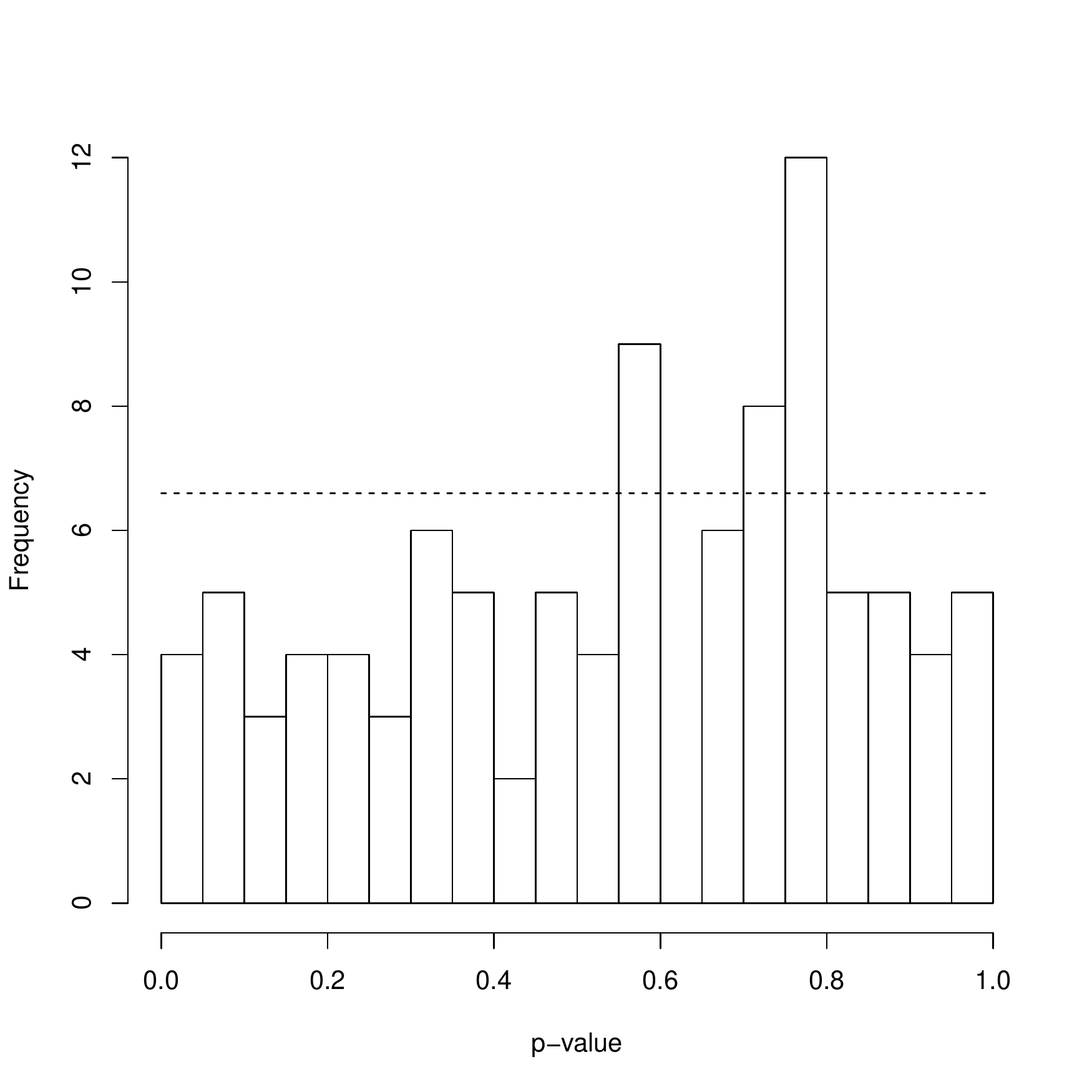}

\caption{\textbf{: Histogram of p-values for migration in simulated data.} We generated 100 tree-like datasets using the topology in Figure 2A in the main text. We then randomly chose two populations (without replacement), added a migration edge between the two populations, and tested for significance using the procedure described in the main text. Plotted is the histogram of p-values for the significance test. If the p-values are properly calibrated, this distribution should be uniform (dotted line). Though the distribution is not completely uniform, there is no skew towards low p-values.}\label{fig_pval}

\end{center}
\end{figure}

\begin{figure}
\begin{center}
\includegraphics[scale = 1]{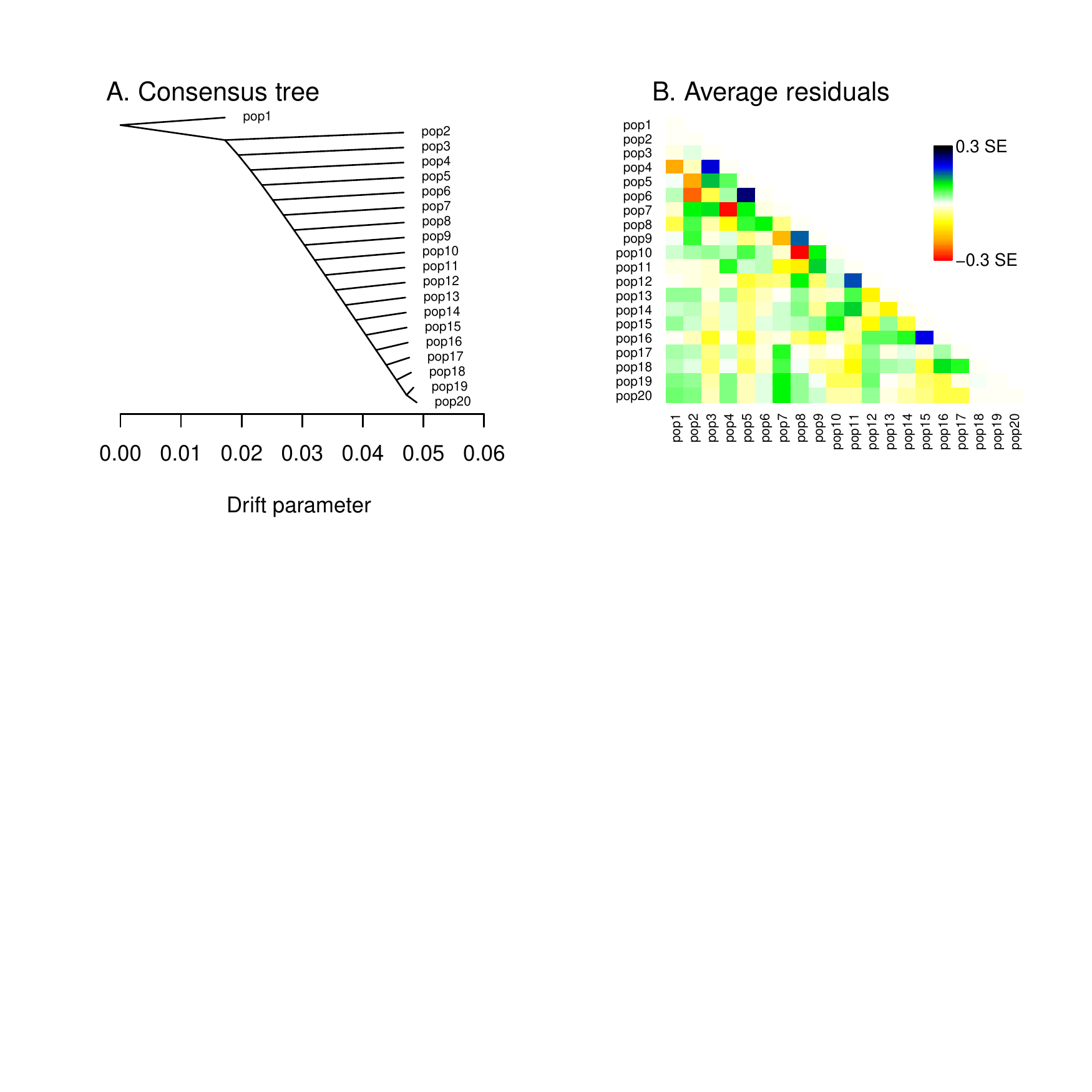}

\caption{\textbf{: Consensus tree in simulations with long branches.} We generated 100 tree-like datasets using the topology in Figure 2A in the main text, multiplying all branch lengths by 50. We then inferred the maximum likelihood tree. \textbf{A.} Plotted are the mean branch lengths from the simulations. All simulations resulted in the same inferred topology. \textbf{B.} In each simulation, we scaled the residuals by the average standard error, then averaged these scaled residuals across simulations. Plotted are the mean scaled residuals across the 100 simulations. The most extreme residuals are not large (around 0.3 standard errors), but tend to be present between closely related populations. }\label{fig_longbranch}

\end{center}
\end{figure}
\begin{figure}
\begin{center}
\includegraphics[scale = 1]{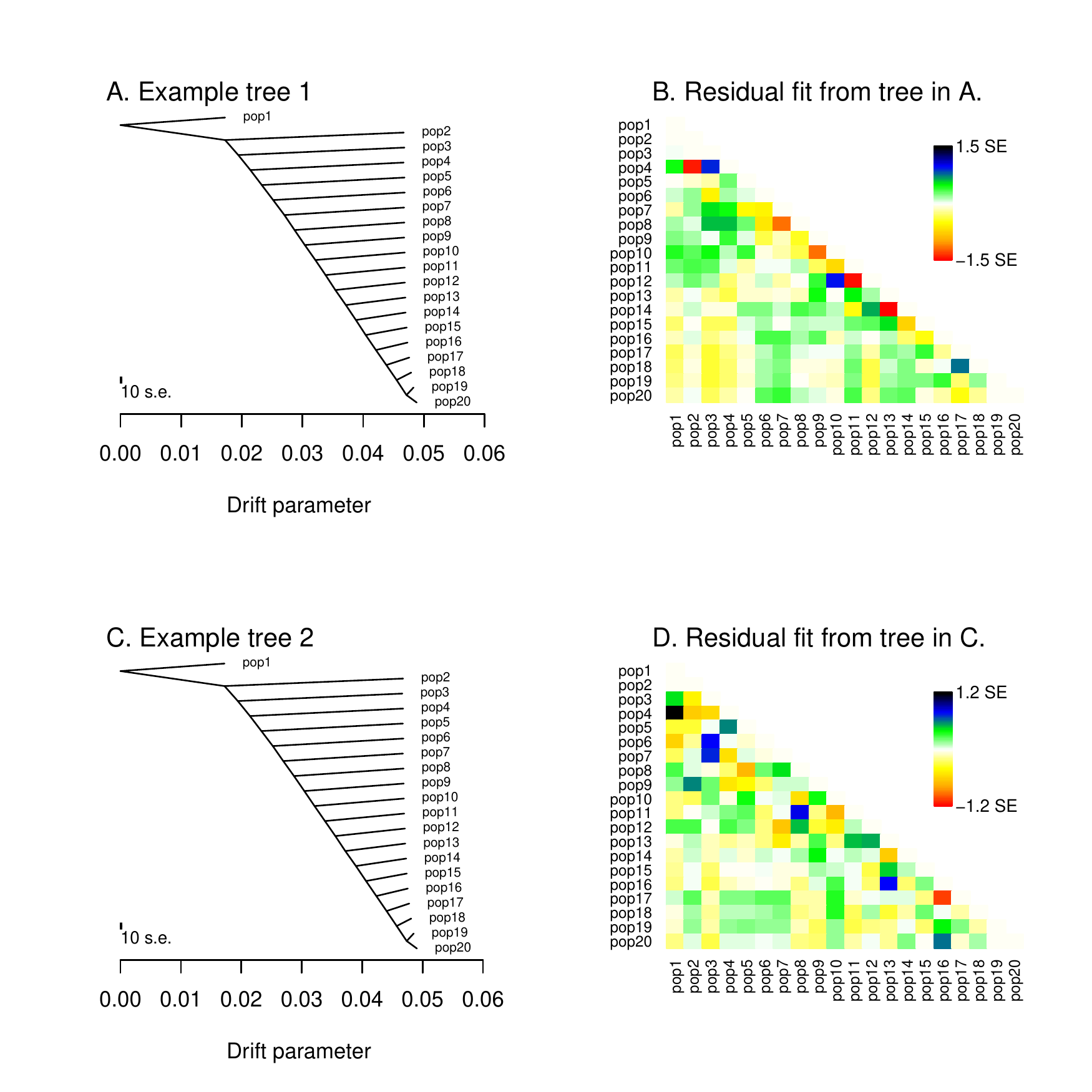}

\caption{\textbf{: Example trees from simulations with long branches.} We generated 100 tree-like datasets using the topology in Figure 2A in the main text, multiplying all branch lengths by 50. We then inferred the maximum likelihood tree. In Supplementary Figure \ref{fig_longbranch}, we show the average inferred tree. Here, we show two representative trees (\textbf{A.} and \textbf{C.} and the residuals corresponding to each tree (\textbf{B.} and \textbf{D.} }\label{fig_longbranch_ex}

\end{center}
\end{figure}

\begin{figure}
\begin{center}
\includegraphics[scale = 1]{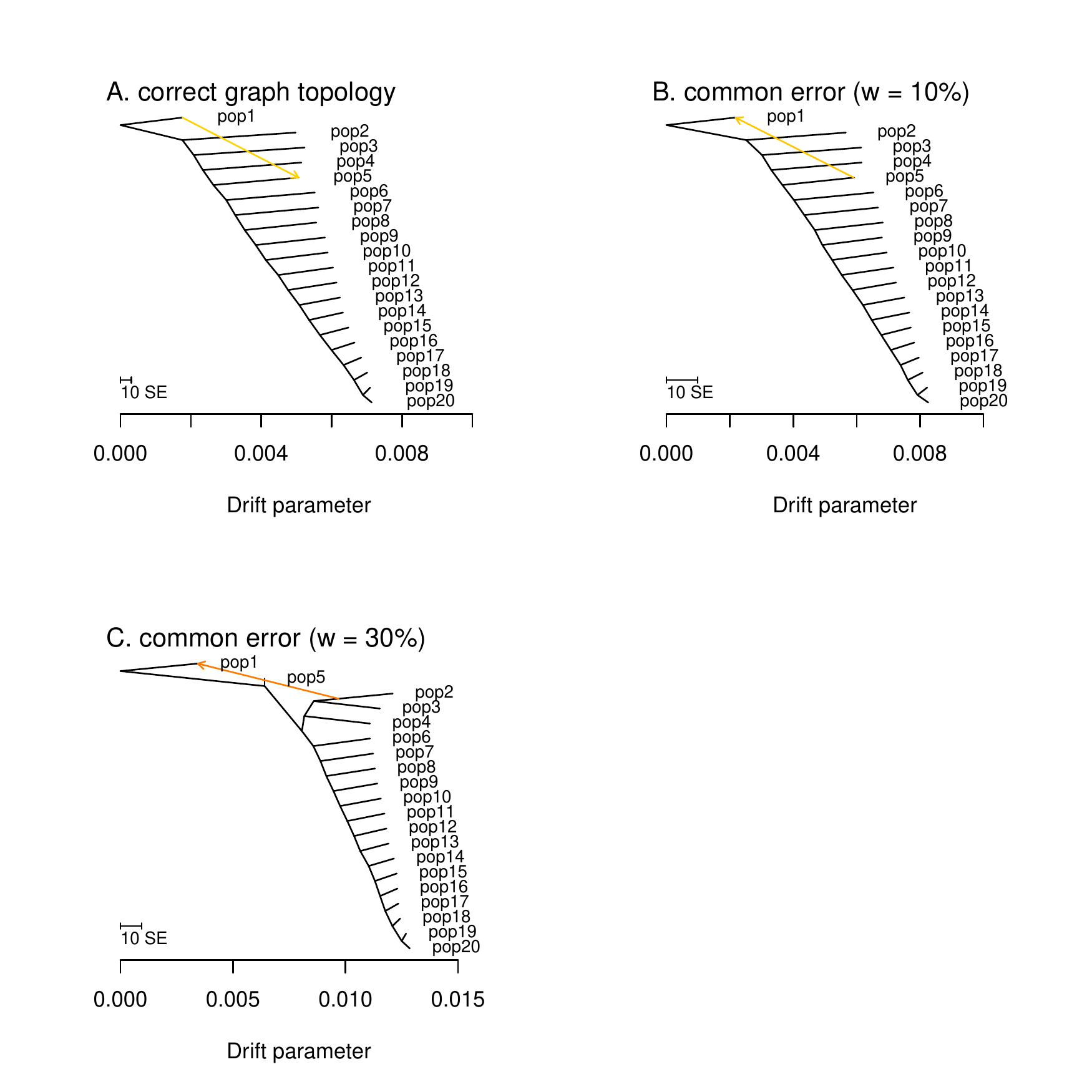}

\caption{\textbf{: Representative errors in simulations.} We examined the simulations in which \emph{TreeMix} did not reach the correct answer. \textbf{A.} The correct topology for the simulations presented in the other panels. \textbf{B.} A representative example of an incorrect topology inferred from the simulations of a migration event with weight 10\% from population 1 to population 5 (this topology accounted for all observed errors). \textbf{C.} A representative example of an incorrect topology inferred from the simulations of a migration event with weight 30\% from population 1 to population 5 (this topology accounted for 95\% of all errors).}\label{fig_error}

\end{center}
\end{figure}

\begin{figure}
\begin{center}
\includegraphics[scale = 0.8]{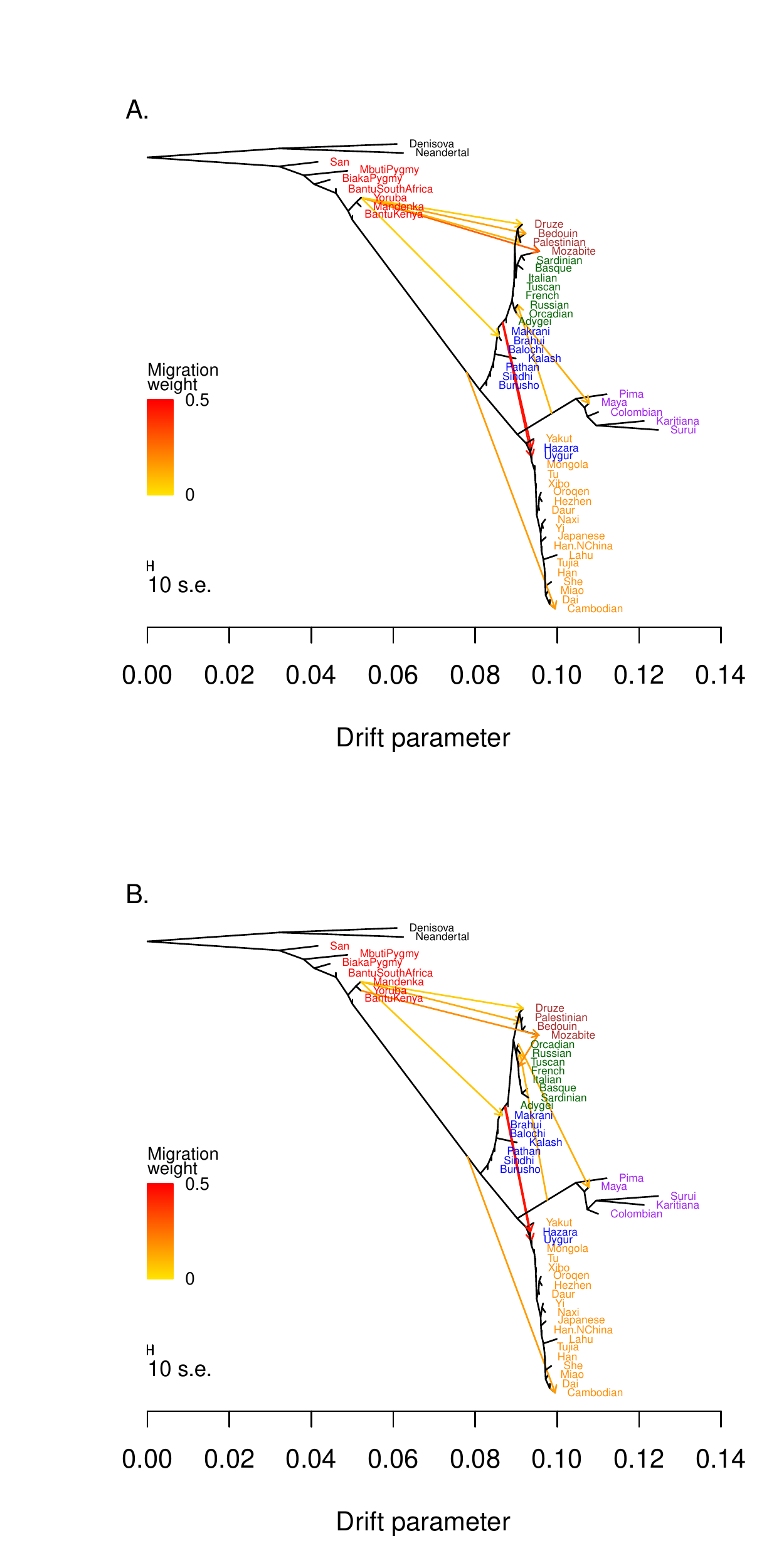}

\caption{\textbf{: Replicate graphs inferred in the human data.} These graphs were generated in an identical manner as Figure 4 in the main text, but using different random input orders for populations during tree-building. All random input orders gave very similar results.}\label{fig_yor}

\end{center}
\end{figure}

\begin{figure}
\begin{center}
\includegraphics[scale = 0.8]{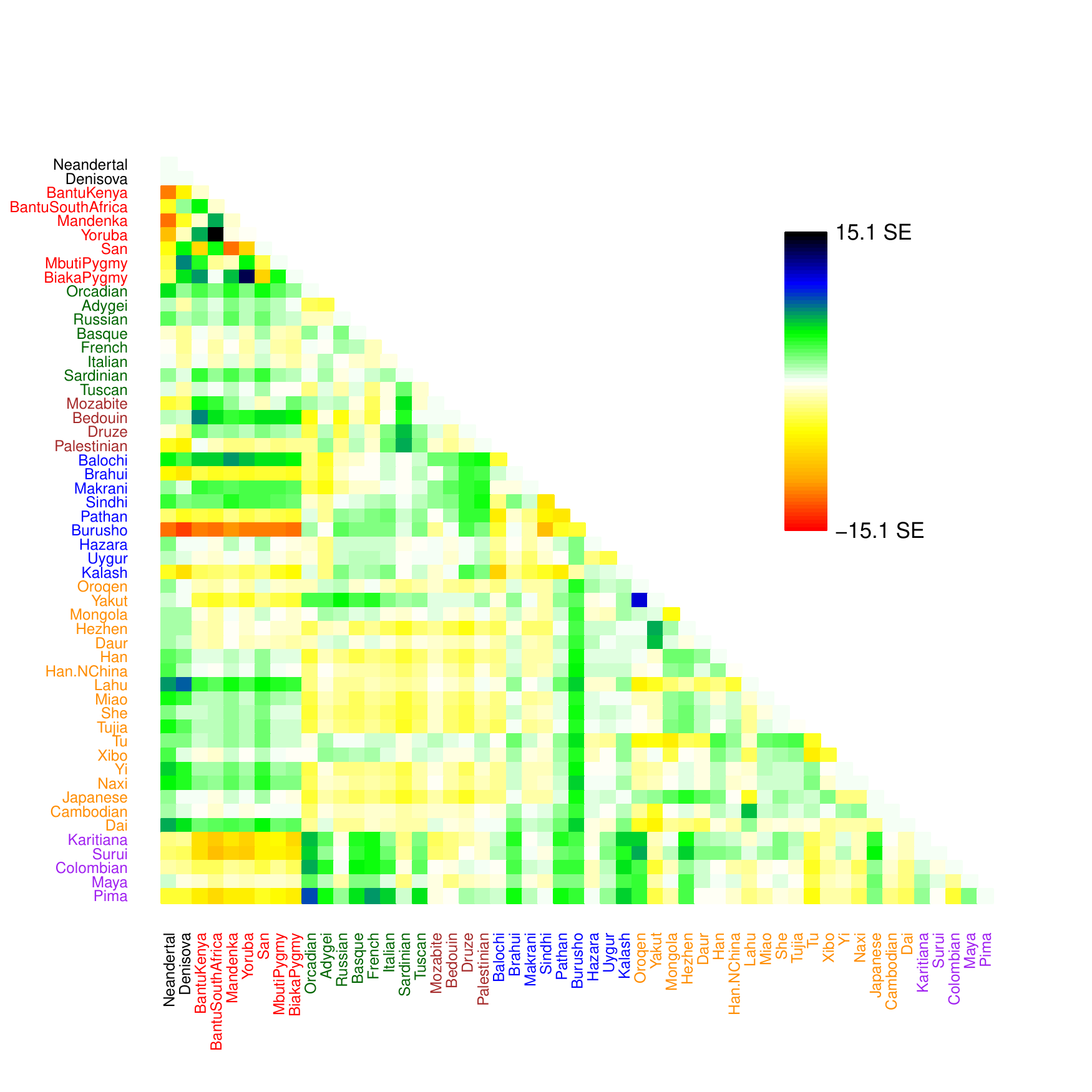}

\caption{\textbf{: Residual fit from graph of human data presented in the main text.} Plotted are the residuals from the fit of the graph presented in Figure 4 in the main text.}\label{fig_yor_resid}

\end{center}
\end{figure}

\begin{figure}
\begin{center}
\includegraphics[scale = 1]{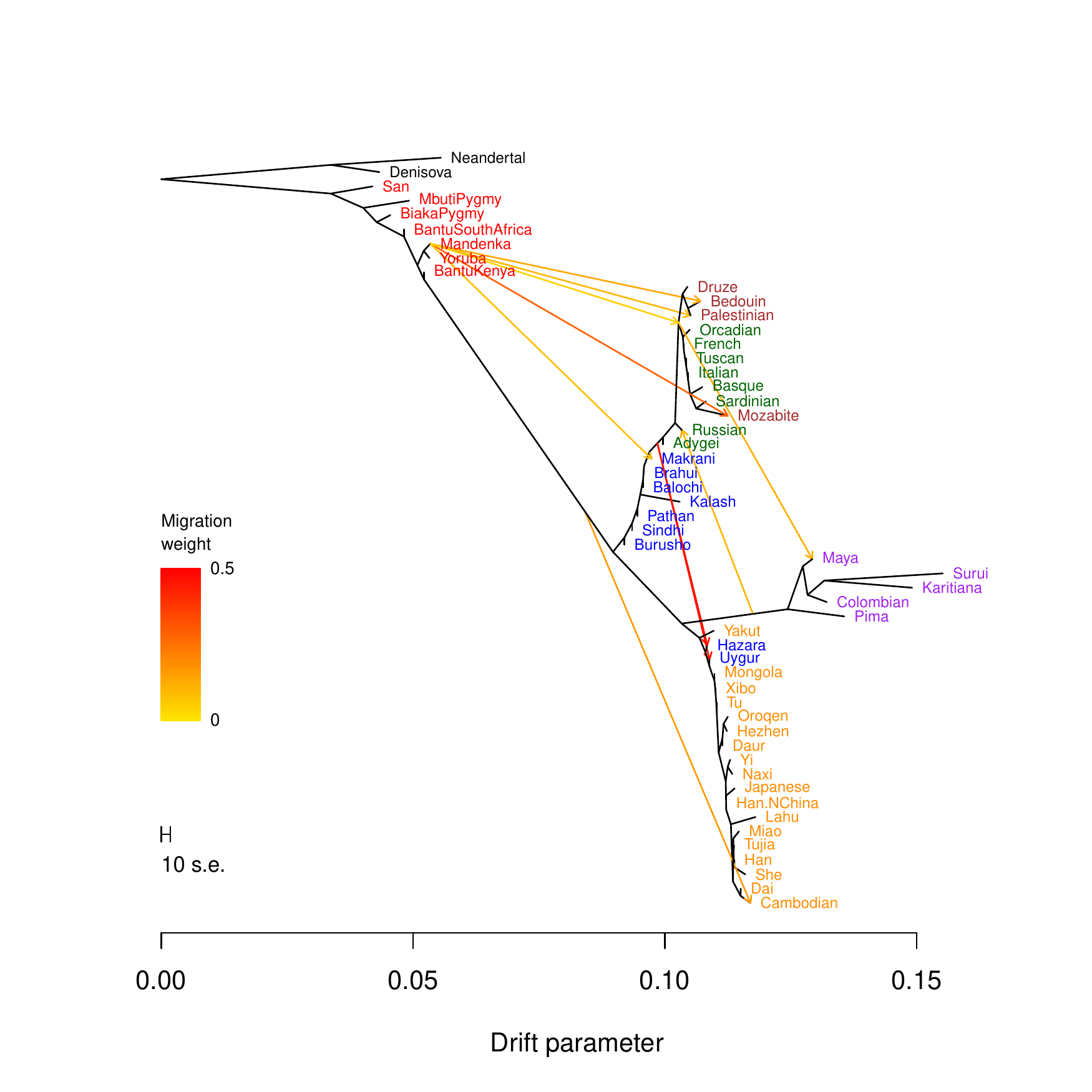}

\caption{\textbf{: Graph inferred from SNPs ascertained in a single French individual.} The graph was generated in an identical manner as Figure 4 in the main text, but using a panel of SNPs ascertained in a single French, rather than a single Yoruban, individual. The inferred graph is extremely similar to that in Figure 4. The one major difference is that, in this graph, the Mozabite appear as an admixture of a Sardinian population rather than a Middle Eastern population; this configuration is seen in some runs of \emph{TreeMix} on the Yoruba-ascertained data (Supplementary Figure \ref{fig_yor}A)}\label{fig_fr}

\end{center}
\end{figure}

\begin{figure}
\begin{center}
\includegraphics[scale = 0.7]{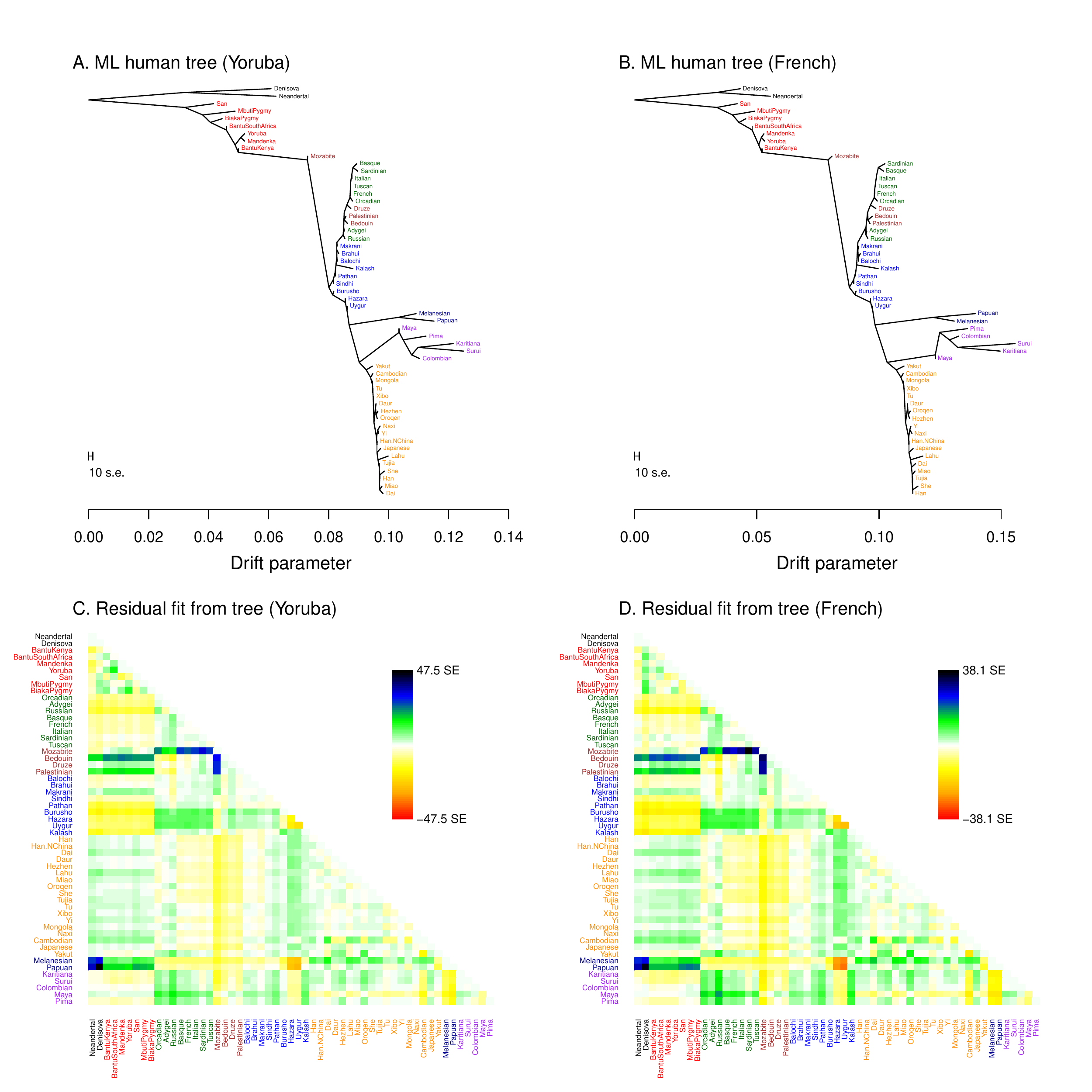}

\caption{\textbf{: Trees inferred using the human data including the Oceanians.} We show the maximum likelihood trees and residuals for the human data including the Oceanian populations, plotted in the same manner as in Figure 3 in the main text. Trees were inferred using the panel of SNPs ascertained in a single Yoruban individual (\textbf{A.} and \textbf{C.}) and the panel of SNPs ascertained in a single French individual (\textbf{B.} and \textbf{D.}). See Supplementary Material for discussion. }\label{fig_woc_tree}

\end{center}
\end{figure}

\begin{figure}
\begin{center}
\includegraphics[scale = 0.8]{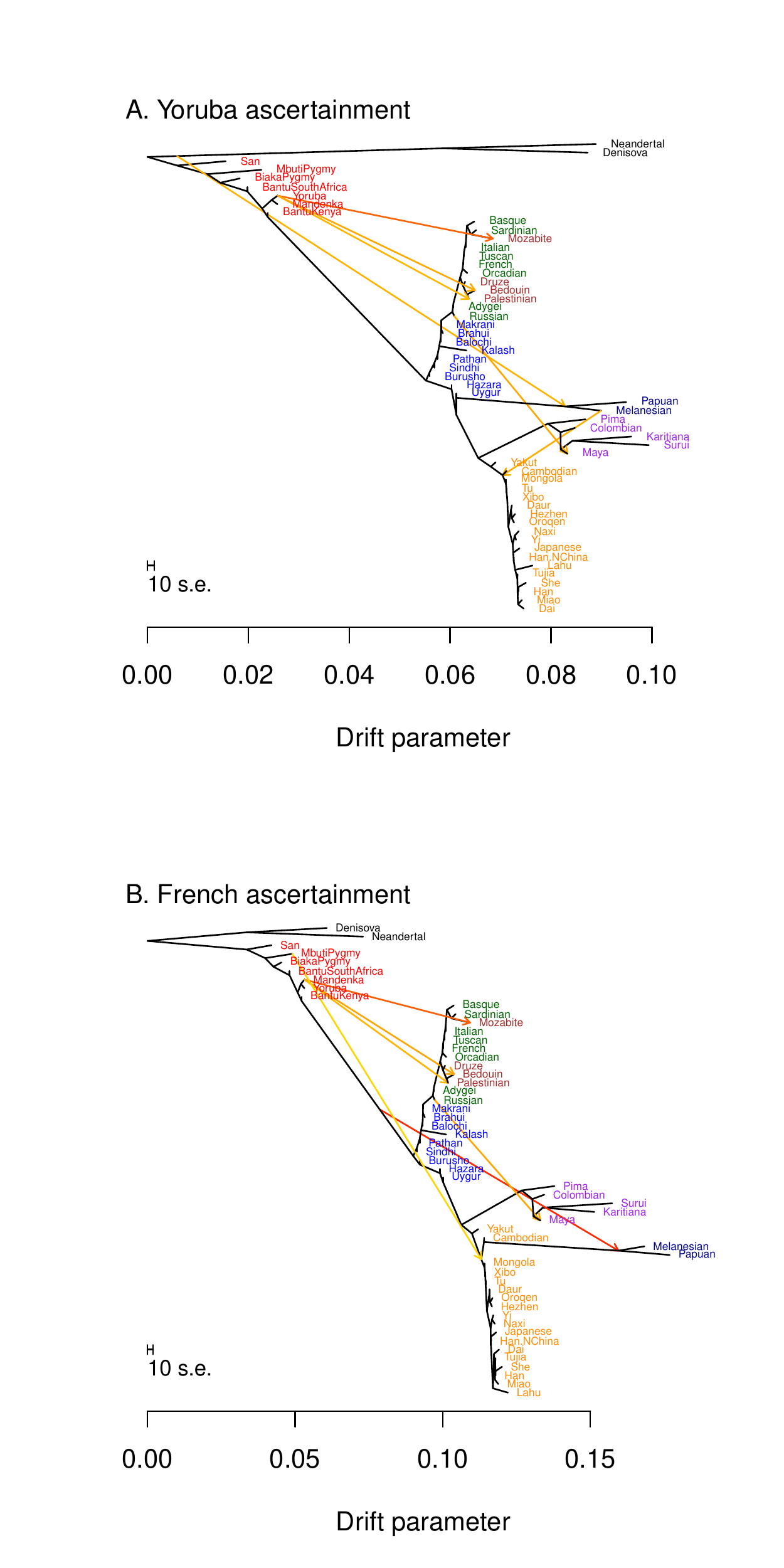}

\caption{\textbf{: Graphs inferred using the human data including the Oceanians.} We show the maximum likelihood graphs for the human data including the Oceanian populations, plotted in the same manner as in Figure 4 in the main text. Six migration edges were inferred in each graph. Graphs were inferred using the panel of SNPs ascertained in a single Yoruban individual (\textbf{A.}) and the panel of SNPs ascertained in a single French individual (\textbf{B.}). See Supplementary Material for discussion.}\label{fig_woc_graphs}

\end{center}
\end{figure}

\begin{figure}
\begin{center}
\includegraphics[scale = 0.8]{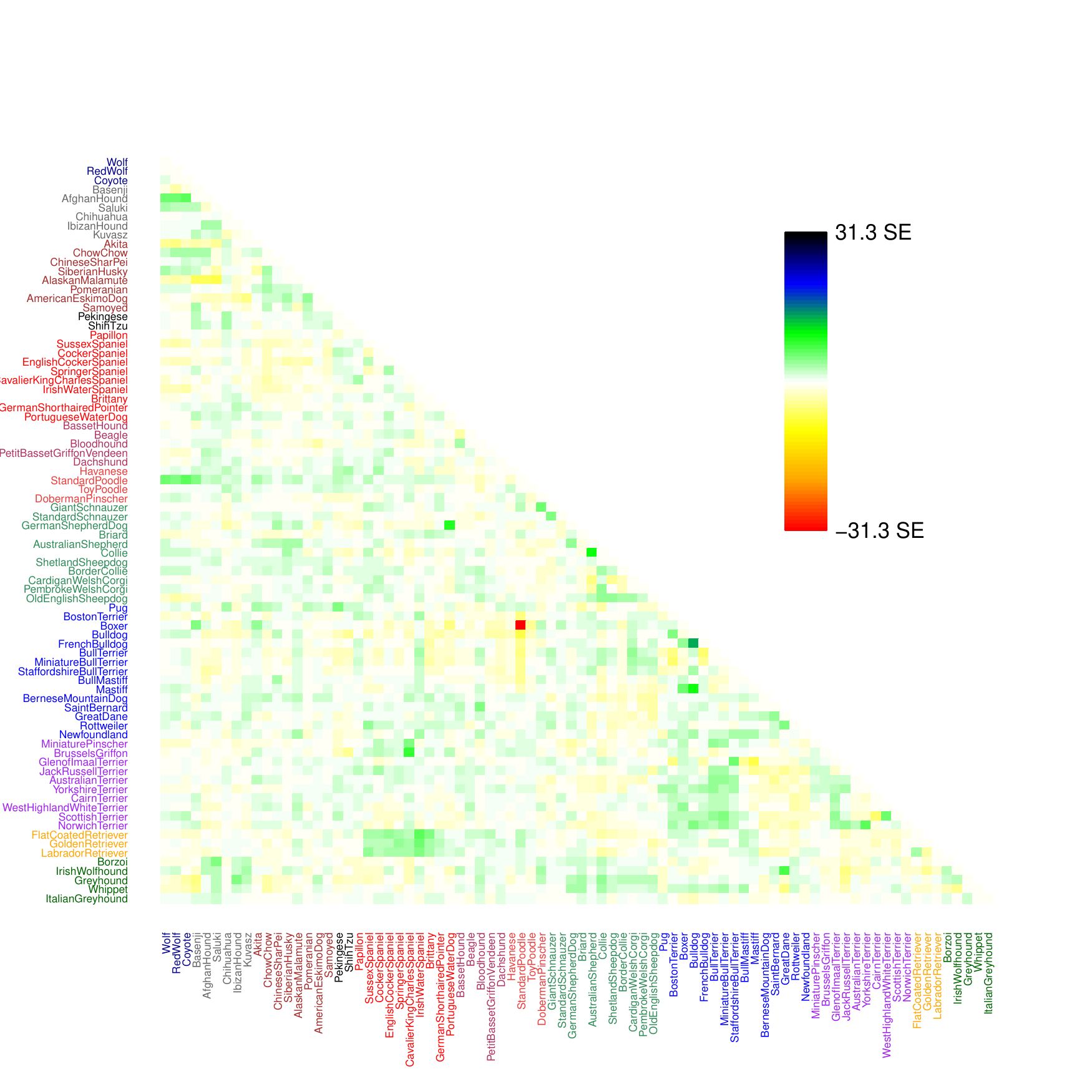}

\caption{\textbf{: Residual fit from graph of dog data presented in the main text.} Plotted are the residuals from the fit of the graph presented in Figure 6 in the main text.}\label{fig_dog_resid}

\end{center}
\end{figure}

\begin{figure}
\begin{center}
\includegraphics[scale = 0.8]{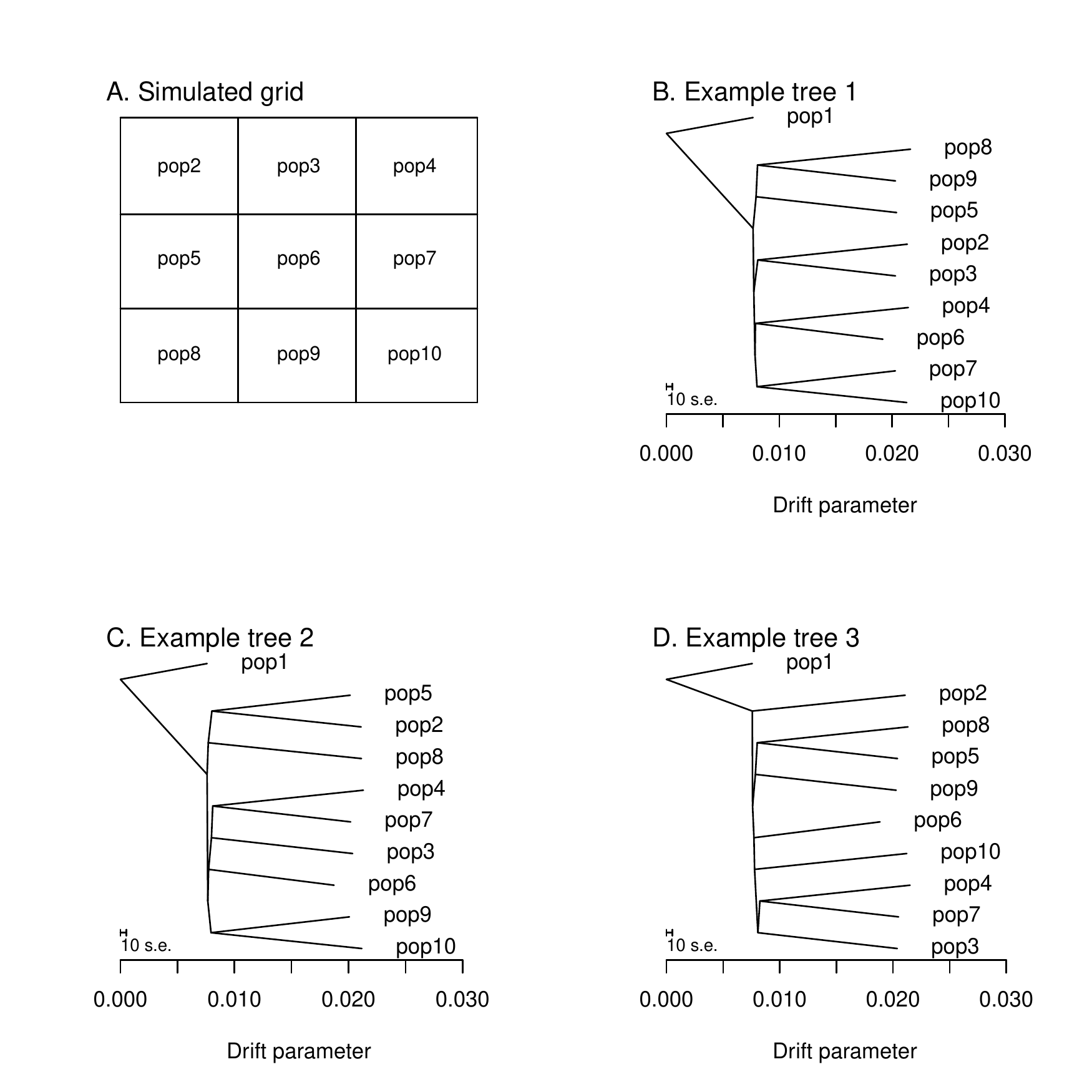}

\caption{\textbf{: \emph{TreeMix} run on populations with continuous migration.} We simulated a set of populations on a lattice, where each population has constant gene flow at a rate of $4Nm = 0.1$ with neighboring populations.  All populations split from an outgroup 16,400 generations in the past. The exact \emph{ms} command is given in the Supplementary Material. The configuration of the lattice is presented in \textbf{A.} (population 1 is the outgroup).  \emph{TreeMix} inferred no consistent tree structure. Three representative trees are presented in \textbf{B.-D.}  }\label{fig_lattice}

\end{center}
\end{figure}

\begin{figure}
\begin{center}
\includegraphics[scale = 0.8]{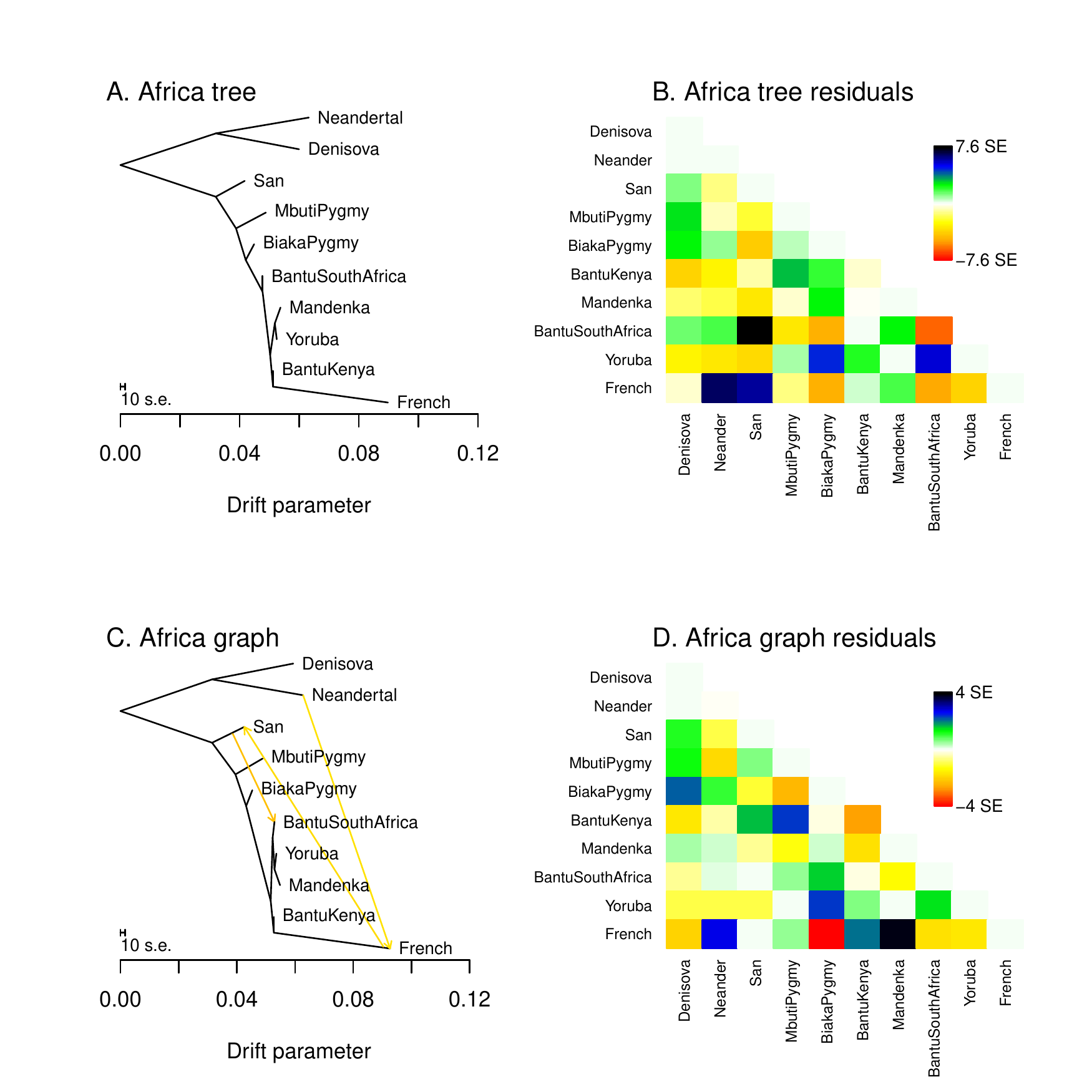}

\caption{\textbf{: \emph{TreeMix} run on human data using only a single non-African population.} We inferred the maximum likelihood tree (\textbf{A.}) using only the African populations and one non-African population (French), using SNPs identified in a single Yoruban individual. In examining the residuals (\textbf{B.}), a relationship between the French and the Neandertal is clear. We then inferred three migration events (\textbf{C.}), where we do see that the French contain some Neandertal ancestry ($w = 1.2\%$). Residual fit for this graph is shown in \textbf{D.}}\label{fig_woc_graphs}

\end{center}
\end{figure}

\clearpage
\bibliography{bib}